\documentclass{article}

\usepackage[nonatbib,preprint]{neurips_2024}

\usepackage[utf8]{inputenc} % allow utf-8 input
\usepackage[T1]{fontenc}    % use 8-bit T1 fonts
\usepackage{hyperref}       % hyperlinks
\usepackage{url}            % simple URL typesetting
\usepackage{booktabs}       % professional-quality tables
\usepackage{amsfonts}       % blackboard math symbols
\usepackage{nicefrac}       % compact symbols for 1/2, etc.
\usepackage{microtype}      % microtypography
\usepackage{xcolor}         % colors
\usepackage{pifont}

\usepackage{multirow}
\usepackage{enumitem}
\usepackage[ruled,linesnumbered]{algorithm2e}
\usepackage{bm}
\usepackage{xspace}
\newcommand{\model}{\textsc{RALLRec+}\xspace}
\usepackage{graphicx}

\newcommand{\priormodel}{\textsc{RALLRec}\xspace}
\usepackage{graphicx}

\usepackage{subfigure}
\usepackage{url}
\usepackage{balance}
\usepackage{amsmath}

\usepackage{hyperref}       % hyperlinks
\hypersetup{colorlinks=true,linkcolor=black,citecolor=black,urlcolor=black}

\definecolor{c1}{HTML}{DC143C}
\definecolor{c2}{HTML}{32CD32}

\title{\model: Retrieval Augmented Large Language
Model Recommendation with Reasoning}

\author{%
  Sichun Luo\textsuperscript{1,2}, Jian Xu\textsuperscript{3}, Xiaojie Zhang\textsuperscript{2}, Linrong Wang\textsuperscript{4}, Sicong Liu\textsuperscript{5}, Hanxu Hou\textsuperscript{1*}, Linqi Song\textsuperscript{2}\thanks{Corresponding Author}
    \\
  \textsuperscript{1}Dongguan University of Technology \quad \textsuperscript{2}City University of Hong Kong \\
  \textsuperscript{3}Tsinghua University \quad
  \textsuperscript{4}Chinese University of Hong Kong \quad
  \textsuperscript{5}Xiamen University \\
  \texttt{sichunluo2@gmail.com} \\
}

\begin{document}

\maketitle

\begin{abstract}
Large Language Models (LLMs) have been integrated into recommender systems to enhance user behavior comprehension. The Retrieval Augmented Generation (RAG) technique is further incorporated into these systems to retrieve more relevant items and improve system performance. However, existing RAG methods have
two shortcomings.
\textit{(i)} In the \textit{retrieval} stage, they rely primarily on textual semantics and often fail to incorporate the most relevant items, thus constraining system effectiveness. 
\textit{(ii)} In the \textit{generation} stage, they lack explicit chain-of-thought reasoning, further limiting their potential.

In this paper, we propose Representation learning and \textbf{R}easoning empowered retrieval-\textbf{A}ugmented \textbf{L}arge \textbf{L}anguage model \textbf{Rec}ommendation (\model). Specifically, 
for the retrieval stage, we prompt LLMs to generate detailed item descriptions and perform joint representation learning, combining textual and collaborative signals extracted from the LLM and recommendation models, respectively. To account for the time-varying nature of user interests, we propose a simple yet effective reranking method to capture preference dynamics. For the generation phase, we first evaluate reasoning LLMs on recommendation tasks, uncovering valuable insights. Then we  introduce knowledge-injected prompting and consistency-based merging approach to integrate reasoning LLMs with general-purpose LLMs, enhancing overall performance. Extensive experiments on three real-world datasets validate our method’s effectiveness. {Code is available at \url{https://github.com/sichunluo/RALLRec_plus}.}
\end{abstract}

\section{Introduction}

% Recommendation systems are widely utilized across various domains \cite{lu2015recommender}. 

Large language models (LLMs) have demonstrated significant potential in many domains due to impressive world knowledge and reasoning capability \cite{achiam2023gpt,openai2024o1,dubey2024llama}.
Recently,  LLMs have been integrated into recommendation tasks \cite{zhao2023recommender,luo2024privacy,luo2024integrating,wu2024survey}. One promising direction for LLM-based recommendations, referred to as LLMRec, involves directly prompting the LLM to perform recommendation tasks in a text-based format \cite{bao2023tallrec,zhang2023chatgpt,luo2024large}.
However, simply using prompts with recent user history can be suboptimal, as they may contain irrelevant information that distracts the LLMs from the task at hand. To address this challenge, ReLLa \cite{lin2024rella} incorporates a retrieval augmentation technique, which retrieves the most relevant items and includes them in the prompt. This approach aims to improve the understanding of the user profile and improve the performance of recommendations. Furthermore, GPT-FedRec \cite{zeng2024federated} proposes a hybrid Retrieval Augmented Generation mechanism to enhance privacy-preserving recommendations by using both an ID retriever and a text retriever.
% addressing data sparsity and heterogeneity 

% \begin{figure}[t]
% \centering
% \includegraphics[width=0.4\textwidth]{p5.png}
% \vspace{-3ex}
%     \caption{\model with embedding, retrieval and reranker.}
% \label{fig:pipeline}
% % \vspace{-6ex}
% \end{figure}

Despite the advancements, current methods have limitations. ReLLa relies primarily on text embeddings for retrieval, which is suboptimal as it overlooks collaborative semantic information from the item side in recommendations. The semantics learned from text are often inadequate as they typically only include titles and limited contextual information. 
% Furthermore, the collaborative semantics from the user side remain underutilized.
GPT-FedRec does not incorporate user's recent interest, and the ID based retriever and text retrieval are in a separate manner, which may not yield the best results.
The integration of text and collaborative information presents challenges as these modalities are not inherently aligned.

Another challenge arises in the generation phase. Existing work typically prompts general-purpose LLMs to generate answers, resulting in models that implicitly map inputs to outputs without explicit reasoning steps. This approach reduces explainability and interpretability, as the chain-of-thought reasoning~\cite{wei2022chain} is overlooked, thereby constraining the model’s potential. Recently, reasoning LLMs, such as the OpenAI-o1 \cite{openai2024o1} and DeepSeek-R1~\cite{guo2025deepseek}, have garnered significant attention for their advanced reasoning capabilities. However, their suitability for recommendation tasks remains unexplored. Although some prior efforts have attempted to integrate reasoning ability into recommendation systems \cite{bismay2024reasoningrec,tsai2024leveraging}, these approaches often rely on specialized workflows that lack adaptability across domains. In contrast, we propose a training-free method to leverage reasoning LLMs, enhancing their flexibility and applicability.

% They use reinformanent learning to train the 671B R1 model, and distill the model to smaller dense model, such as 

In this work, we propose Representation Learning and \underline{R}easoning enhanced Retrieval-\underline{A}ugmented \underline{L}arge \underline{L}anguage Models for \underline{Rec}ommendation (\model). 
% Our objective is to enhance the performance of retrieval-augmented LLM recommendations through improved representation learning and reasoning.
Specifically, regarding the retrieval stage, instead of solely relying on abbreviated item titles to extract item representations, we prompt the LLM to generate detailed descriptions for items utilizing its world knowledge. These generated descriptions are used to extract improved item representations. This representation is concatenated with the abbreviated item representation.
Subsequently, we obtain collaborative semantics for items using a recommendation model. This collaborative semantic is aligned with textual semantics through self-supervised learning to produce the final representation. This enhanced representation is used to retrieve items, thereby improving Retrieval-Augmented Large Language Model recommendations.
To enhance the generation stage, we first evaluate the reasoning LLM on recommendation tasks, uncovering intriguing insights. Based on these findings, we propose an effective knowledge-injected prompting method. By incorporating prior knowledge from recommendation experts, this approach enables the reasoning LLM to deliver more precise predictions. Additionally, we introduce a consistency-based merging technique to integrate the reasoning LLM with a general-purpose LLM, further improving overall performance.

Note that some preliminary findings were reported at the ACM Web Conference 2025 (WWW'25) \cite{xu2025rallrec}. The enhancements made in this extended version of the work are as follows:
\begin{itemize}[left=0em]
\item
We upgrade the \priormodel framework to \model. While \priormodel solely enhances the retrieval stage through representation learning, \model extends this by focusing on the generation stage. Specifically, we evaluate reasoning LLM on recommendation tasks and propose simple yet effective strategies to boost model performance.
\item
We expand our experimental evaluation by incorporating additional models and settings, providing clearer evidence of \model’s superiority.
\item
Lastly, we restructure the paper to better articulate the motivations, objectives, and advancements of these revisions, offering readers deeper insight into this extended work.
\end{itemize}

In a nutshell, our contribution is threefold.
\vspace{-1ex}
\begin{itemize}[left=0em]
    \item We propose \model, which incorporates collaborative information and learns joint representations to retrieve more relevant items, thereby enhancing the retrieval-augmented large language model recommendation.

    \item We evaluate reasoning models on recommendation tasks, uncovering several interesting insights. Leveraging these findings, we propose a simple yet effective framework to adapt reasoning models into existing retrieval-augmented LLM recommendation systems.
    
    \item We conduct extensive experiments to demonstrate the effectiveness of our proposed \model, further revealing valuable findings.

\end{itemize}

    % \item We utilize contrastive learning to align text semantics with collaborative semantics, leveraging user-side collaborative information to improve model performance.

    % \item We design a novel reranker that takes into account both the semantic similarity to the target item and the timestamps for boosting the validness of RAG.

% \section{Related work}

% \subsection{Large Language Model Recommendation}

% \subsection{Representation Learning in Recommendation}
% In recommendation, MF learns collaborative representations. Subsequently, a graph structure is employed to model more complex and higher-order collaborative representations. Contrastive learning is integrated into the recommendation process by using contrastive loss to develop more uniform representations.
% More recently, LLMs have been incorporated into representation learning. Concurrent with our work, [XXX] aligns collaborative and textual information. However, these efforts do not specifically target Retrieval-Augmented Large Language Model recommendations.

\section{Related Work}

\subsection{LLM for Recommendation}
Recent advancements in LLMs have significantly reshaped recommender systems by leveraging their natural language understanding and generation capabilities for enhanced personalization \cite{luo2024recranker,wu2024survey,luo2024large}. Existing research categorizes LLM-based approaches into two paradigms: discriminative and generative \cite{wu2024survey}. In the discriminative paradigm, LLMs are used to extract textual features, such as user and item embeddings, for traditional recommendation algorithms~\cite{kim2024large}. Conversely, the generative paradigm employs LLMs, such as ChatGPT and Llama, to directly generate recommendations or explanations, excelling in zero-shot and few-shot scenarios~\cite{ji2024genrec}. 

Despite these advancements, simply using prompts with recent user history can be suboptimal, as they may contain
irrelevant information that distracts the LLMs from the task at hand. Thus, retrieval-augmented generation technique is further integrated for better performance.

% For instance, Wang et al. proposed a hierarchical system using LLMs to generate novel interest clusters at the conceptual level, while grounding them with transformer-based sequence recommenders for item-level execution, achieving 32\% improvement in exploration efficiency on industrial platforms . Multimodal extensions like MLLM4Rec further enhance accuracy by integrating visual data with hybrid prompt learning . However, challenges persist in computational efficiency, with methods like prompt distillation  and quantization-aware fine-tuning  being proposed to address latency constraints.

\subsection{Retrieval-Augmented Generation}

% Retrieval-Augmented Generation (RAG) enhances large language model (LLM) performance by dynamically incorporating relevant information from external knowledge resources during inference \cite{lewis2020retrieval,gao2023retrieval}. This paradigm has demonstrated success across diverse NLP tasks, including language modeling \cite{guu2020retrieval}, open-domain question answering \cite{izacard2021distilling}, and knowledge-intensive dialogue systems \cite{DBLP:journals/corr/abs-2107-07566}. 
% ReLLa \cite{lin2024rella} and GPT-FedRec \cite{zeng2024federated} intergarted RAG into recommendation.

% However, exsiting method do not learn good representation and could lead to sub optimal results.

Retrieval-Augmented Generation (RAG) improves LLM performance by dynamically integrating relevant external knowledge during inference~\cite{lewis2020retrieval,gao2023retrieval,asai2023self}. This approach has proven effective across various NLP tasks, such as language modeling~\cite{guu2020retrieval}, open-domain question answering~\cite{izacard2021distilling}, and knowledge-intensive dialogue systems~\cite{DBLP:journals/corr/abs-2107-07566}. In the recommendation domain, RAG has been adopted to enhance item retrieval and user understanding.
Notably, ReLLa~\cite{lin2024rella} leverages RAG to augment LLMs with retrieved textual semantics for improved user comprehension and recommendation performance. Similarly, GPT-FedRec~\cite{zeng2024federated} employs RAG within a federated learning framework to ensure privacy-preserving recommendations. 

However, existing methods often fail to learn comprehensive embeddings, leading to suboptimal retrieval and generation results. Our work addresses this limitation by introducing joint representation learning and reasoning enhancements.

% Recent work has explored hybrid architectures that combine dense retrieval with parametric memory \cite{wang2023knowledge}, though challenges remain in optimizing retrieval latency and grounding hallucinations. Our method builds on these insights while introducing a novel iterative retrieval-reasoning mechanism to address cascading errors in multi-hop scenarios.

\subsection{LLM Reasoning}

The reasoning capabilities of LLMs have evolved significantly. Early work on Chain-of-Thought (CoT) prompting , introduced by~\cite{wei2022chain}, demonstrated that prompting LLMs to generate step-by-step reasoning traces improves performance on complex tasks such as arithmetic and symbolic reasoning. Building on this, learning-based methods have sought to embed reasoning capabilities directly into LLMs, reducing reliance on external prompts. STaR~\cite{zelikman2022star} iteratively fine-tunes models on self-generated reasoning traces, baking CoT-like behavior into the model itself. Similarly,~\cite{lightman2023let} trains process reward models (PRMs) to evaluate intermediate reasoning steps, outperforming outcome-based rewards in tasks requiring multi-step logic. OpenAI o1 model~\cite{openai2024o1} marks a significant leap in this direction, leveraging large-scale reinforcement learning (RL) to natively integrate CoT during inference. Unlike traditional autoregressive LLMs, o1 employs test-time compute to iteratively refine its reasoning, achieving state-of-the-art results on challenging benchmarks.
% Recent efforts like DeepSeek-R1 \cite{deepseek2024r1} combine supervised fine-tuning (SFT) with a progress-aware reward model to align reasoning trajectories with human preferences. However, these approaches remain constrained by static parametric knowledge. Our work bridges this gap by integrating dynamic retrieval with curriculum-based reinforcement learning, enabling continuous knowledge refinement during reasoning.

However, none of the previous works attempted to apply o1-like reasoning to LLM recommendation, resulting in a research gap. In this paper, we aim to bridge this gap by evaluating and leveraging the zero-shot reasoning ability of LLMs.

\section{Evaluation}
\label{sec:analysis}

\begin{figure}[t]
\centering
\includegraphics[width=0.89\textwidth]{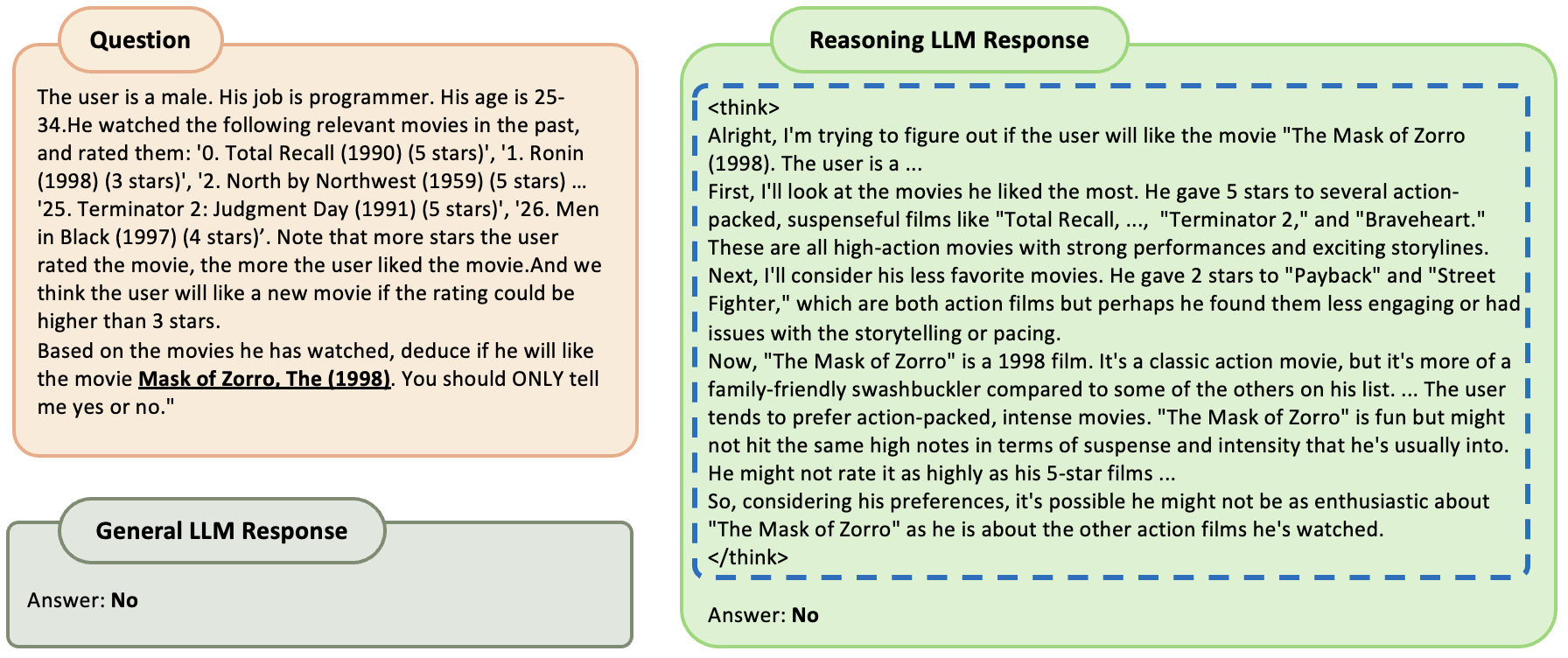}
    \caption{Example of the response generated by general LLM and reasoning LLM.}
\label{fig:example}
% \vspace{-3ex}
\end{figure}

We first conduct an evaluation of reasoning LLM on recommendation tasks and yield some interesting findings. An example of response generated by general LLM and reasoning LLM is shown in Figure \ref{fig:example}.

\textbf{Dataset.}
In this paper, we focus on the click-through rate (CTR) prediction \cite{lin2024rella}. 
We utilize three widely used public datasets: BookCrossing \cite{ziegler2005improving}, MovieLens \cite{harper2015movielens},  and Amazon \cite{ni2019justifying}.
For the MovieLens dataset, we select the MovieLens-1M subset, and for the Amazon dataset, we focus on the Movies \& TV subset. We apply the 5-core strategy to filter out long-tailed users/items with less than 5 records. The statistics are shown in Table~\ref{tab:datasets}.

\begin{table}[!t]
    \caption{The dataset statistics.}
    % \vspace{-8pt}
    \centering
    \resizebox{0.6\textwidth}{!}{
    \renewcommand\arraystretch{1.1}
    \begin{tabular}{c|cccccc}
    \toprule
     Dataset   & \#Users & \#Items & \#Samples & \#Fields & \#Features \\ 
     \midrule
     BookCrossing  & 8,723 & 3,547 & 227,735 & 10 & 14,279 \\
     MovieLens & 6,040 & 3,952 & 970,009 & 9 & 15,905 \\
     Amazon & 14,386 & 5,000 & 141,829 & 6 & 22,387 \\ 
     \bottomrule
    \end{tabular}
    }
    % \vspace{-5pt}
    \label{tab:datasets}
\end{table}

\begin{table*}[!t]
\caption{The comparison of general purpose model and reasoning model. The best results are highlighted in boldface.
}
% \vspace{-8pt}
\label{tab:compare_reason_llm}
\resizebox{\textwidth}{!}{
\renewcommand\arraystretch{1.0}
\begin{tabular}{c|ccc|ccc|ccc}
\toprule
\multicolumn{1}{c|}{\multirow{2}{*}{Model}} & \multicolumn{3}{c|}{BookCrossing} & \multicolumn{3}{c|}{MovieLens} & \multicolumn{3}{c}{Amazon} \\ 
\multicolumn{1}{c|}{} & AUC $\uparrow$  & Log Loss $\downarrow$& ACC $\uparrow$ & AUC $\uparrow$  & Log Loss $\downarrow$& ACC $\uparrow$ & AUC $\uparrow$  & Log Loss $\downarrow$& ACC $\uparrow$ \\ 
   \midrule 
Llama-3.1-8B-Instruct & 0.5894 & \textbf{0.6839} & 0.5418 & 0.5865 & 0.6853 & 0.5591 & \textbf{0.7025} & 0.7305 & 0.4719 \\ 
DeepSeek-R1-Distill-Llama-8B & \textbf{0.6147} & 0.7065 & \textbf{0.5487} & \textbf{0.5944} & \textbf{0.6850} & \textbf{0.5752} & 0.6874 & \textbf{0.5392} & \textbf{0.7792} \\
\textit{Improvement (\%)} & \textcolor{c1}{\textbf{+4.29}} & \textcolor{c2}{\textbf{-3.30}} & \textcolor{c1}{\textbf{+1.27}} & \textcolor{c1}{\textbf{+1.35}} & \textcolor{c1}{\textbf{+0.04}} & \textcolor{c1}{\textbf{+2.88}} & \textcolor{c2}{\textbf{-2.15}} & \textcolor{c1}{\textbf{+26.19}} & \textcolor{c1}{\textbf{+65.12}} \\
\bottomrule          
\end{tabular}
}

% \vspace{-3ex}
\end{table*}

% We first evaluate reasoning LLM in the recommenadtion task. We aim to answer the following research questions.

% \begin{itemize}[left=0em]
%     \item \textbf{RQ1:}
%     How does reasoning LLMs compare with general purpose LLMs?
%     % How does our proposed \model framework compare with both the conventional recommendation models and the state-of-the-art LLM-based RAG recommendation methods?  
%     \item \textbf{RQ2:} 
%     Will the genrated kengtrh for reasonoig LLMs influence the model performance?
%     % Do the designed components of our model, including the representation learning and alignment, embedding mixture and re-ranking module, function effectively?  
%     % \item \textbf{RQ3:} How do different hyper-parameters and training/inference settings affect the final recommendation performance?  
% \end{itemize}

\textbf{Comparison between Reasoning LLM and General LLM.}
To evaluate the impact of reasoning capabilities in LLMs for recommendation tasks, we compare DeepSeek-R1-Distill-Llama-8B, a model supervised fine-tuned with distilled CoT reasoning data \cite{guo2025deepseek}, against Llama-3.1-8B-Instruct, a general-purpose LLM of the same size. Both models are evaluated using a simple retrieval approach, ensuring a fair comparison. Following the official guidance, we set the temperature to 0.6 for DeepSeek-R1-Distill-Llama-8B models\footnote{\url{https://huggingface.co/deepseek-ai/DeepSeek-R1}}. We run the experiment five times and calculate the average.

The results of this comparison are presented in Table \ref{tab:compare_reason_llm}, where we report the accuracy, log loss, and AUC scores. We observe that the reasoning LLM always achieves better accuracy compared to the general LLM. This improvement may be attributed to the CoT reasoning, which enables the model to follow longer and more accurate reasoning paths, thereby enhancing its decision-making process.
However, we also note that the reasoning LLM exhibits a lower AUC or Log Loss score in some cases. This could be due to the model being overly confident in its predictions, which might lead to a higher rate of false positives or false negatives, adversely affecting these metrics.

\textbf{\textsc{Takeaway} I}: Reasoning LLMs generally outperform general LLMs in recommendation tasks.

\begin{table*}[t]
\caption{The comparison of reasoning model \textit{w/} and \textit{w/o} retrieval. The best results are highlighted in boldface.}
\label{tab:retrieval}
\resizebox{\textwidth}{!}{
\renewcommand\arraystretch{1.0}
\begin{tabular}{c|ccc|ccc|ccc}
\toprule
\multirow{2}{*}{{\textit{w/} retrieval}} & \multicolumn{3}{c|}{BookCrossing} & \multicolumn{3}{c|}{MovieLens} & \multicolumn{3}{c}{Amazon} \\
% \cmidrule(lr){2-4} \cmidrule(lr){5-7} \cmidrule(lr){8-10}
 & AUC $\uparrow$ & Log Loss $\downarrow$ & ACC $\uparrow$ & AUC $\uparrow$ & Log Loss $\downarrow$ & ACC $\uparrow$ & AUC $\uparrow$ & Log Loss $\downarrow$ & ACC $\uparrow$ \\
\midrule
\ding{55} & 0.6147 & 0.7065 & {0.5487} & 0.5944 & 0.6851 & 0.5752 & 0.6874 & 0.5392 & 0.7792 \\
\ding{51} & \textbf{0.6274} & \textbf{0.7064} & \textbf{0.5498} & \textbf{0.6028} & \textbf{0.6805} & \textbf{0.5790} & \textbf{0.7014} & \textbf{0.5335} & \textbf{0.7879} \\
% \m
\textit{Improvement (\%)} & \textcolor{c1}{\textbf{+2.07}} & \textcolor{c1}{\textbf{+0.01}} & \textcolor{c1}{\textbf{+0.20}} & \textcolor{c1}{\textbf{+1.41}} & \textcolor{c1}{\textbf{+0.67}} & \textcolor{c1}{\textbf{+0.66}} & \textcolor{c1}{\textbf{+2.04}} & \textcolor{c1}{\textbf{+1.06}} & \textcolor{c1}{\textbf{+1.12}} \\
\bottomrule
\end{tabular}
}
\end{table*}

\textbf{Analysis of Retrieval in Reasoning LLM.}
We further investigate the impact of retrieval on the performance of reasoning LLMs. Specifically, we employ the representation learning enhanced retrieval mechanism in Sec.~\ref{sec:retrieval} to augment the model's input with relevant information. For a fair comparison, we use the DeepSeek-R1-Distill-Llama-8B model as the base model in all experiments. The results, presented in Table \ref{tab:retrieval}, demonstrate that incorporating retrieval leads to improved model performance across all three datasets. This enhancement can be attributed to the inclusion of more relevant items, which facilitates the model's reasoning process.

\textbf{\textsc{Takeaway II}}: Retrieval augments the model's performance by providing relevant contextual information that enhances reasoning capabilities.

\textbf{Analysis of Response Length in Reasoning LLM.}
We also explore the relationship between the length of the reasoning LLM's response and its performance on recommendation tasks. Prior
 studies in mathematical reasoning have shown that longer responses often lead to better performance \cite{guo2025deepseek}. 
This raises a natural question: \textit{does response length affect LLM recommendation performance?}

\begin{figure}[!t]
    \centering
    % First subplot
    \subfigure[BookCrossing]{
        \includegraphics[width=0.31\columnwidth]{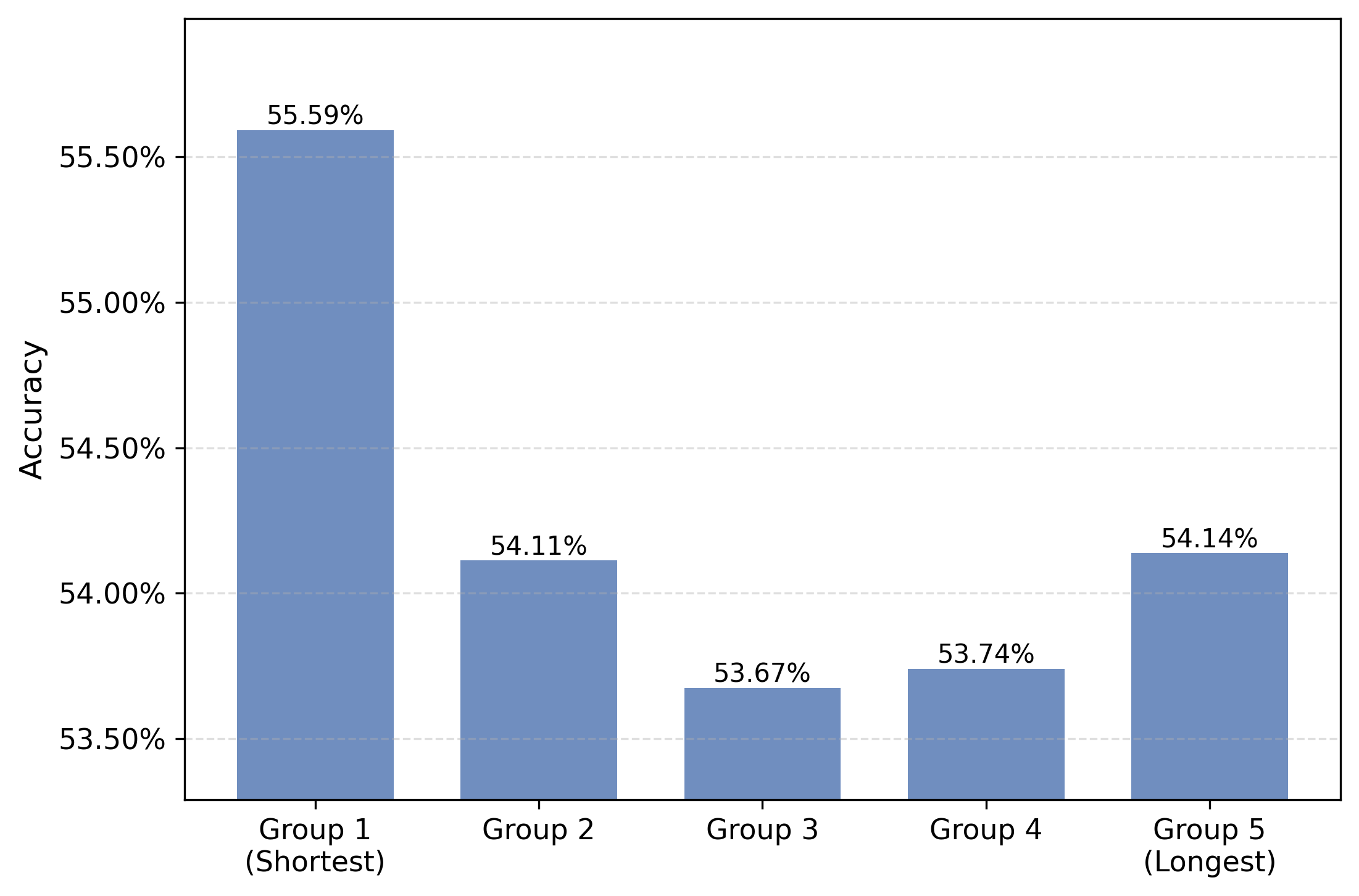} % Replace with your image file
    }
    \hfill
    % Second subplot
    \subfigure[MovieLens]{
        \includegraphics[width=0.31\columnwidth]{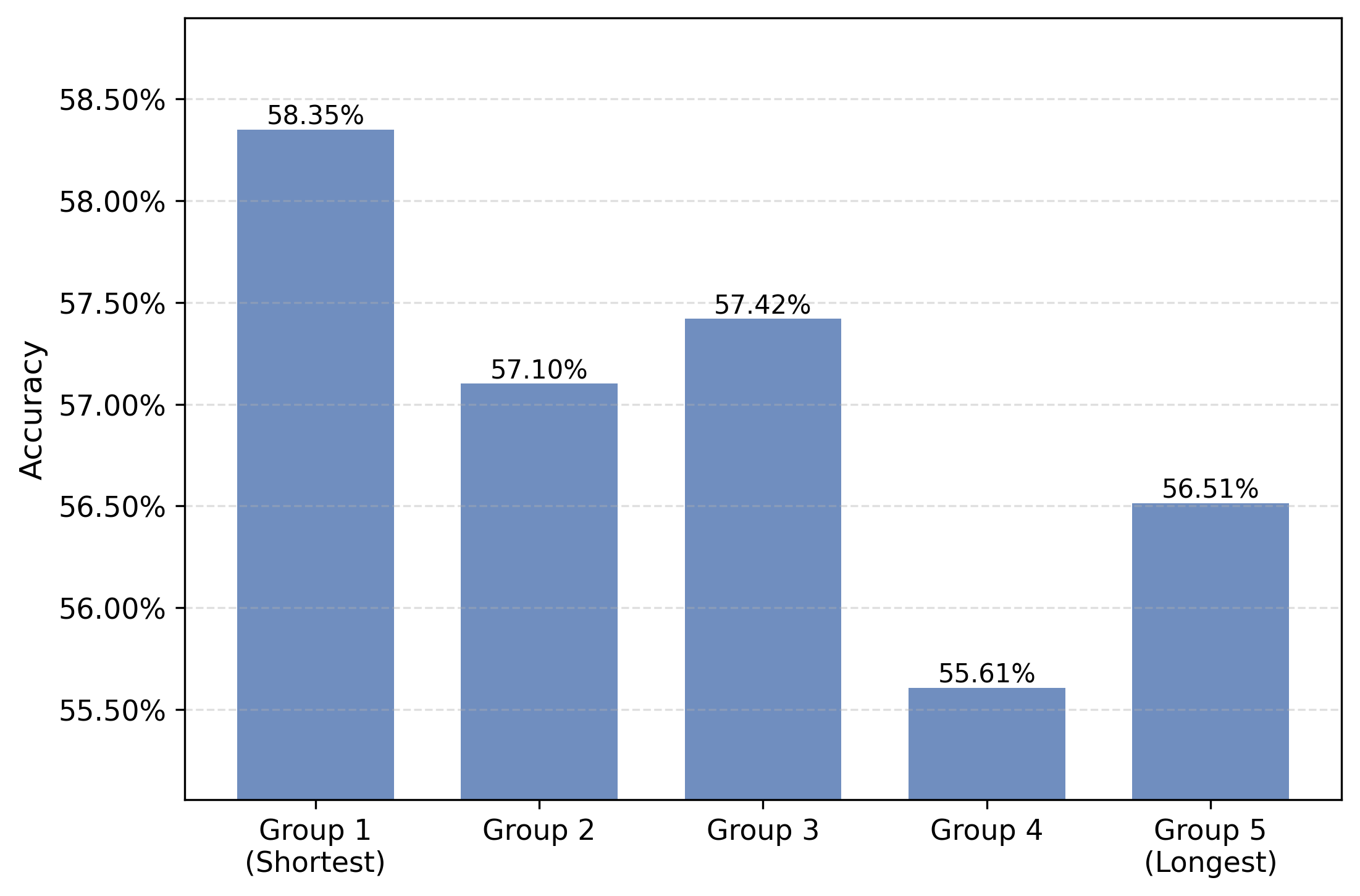} % Replace with your image file
        % \label{fig:subplot2}
        }
    \hfill
    % Second subplot
    \subfigure[Amazon]{
        \includegraphics[width=0.31\columnwidth]{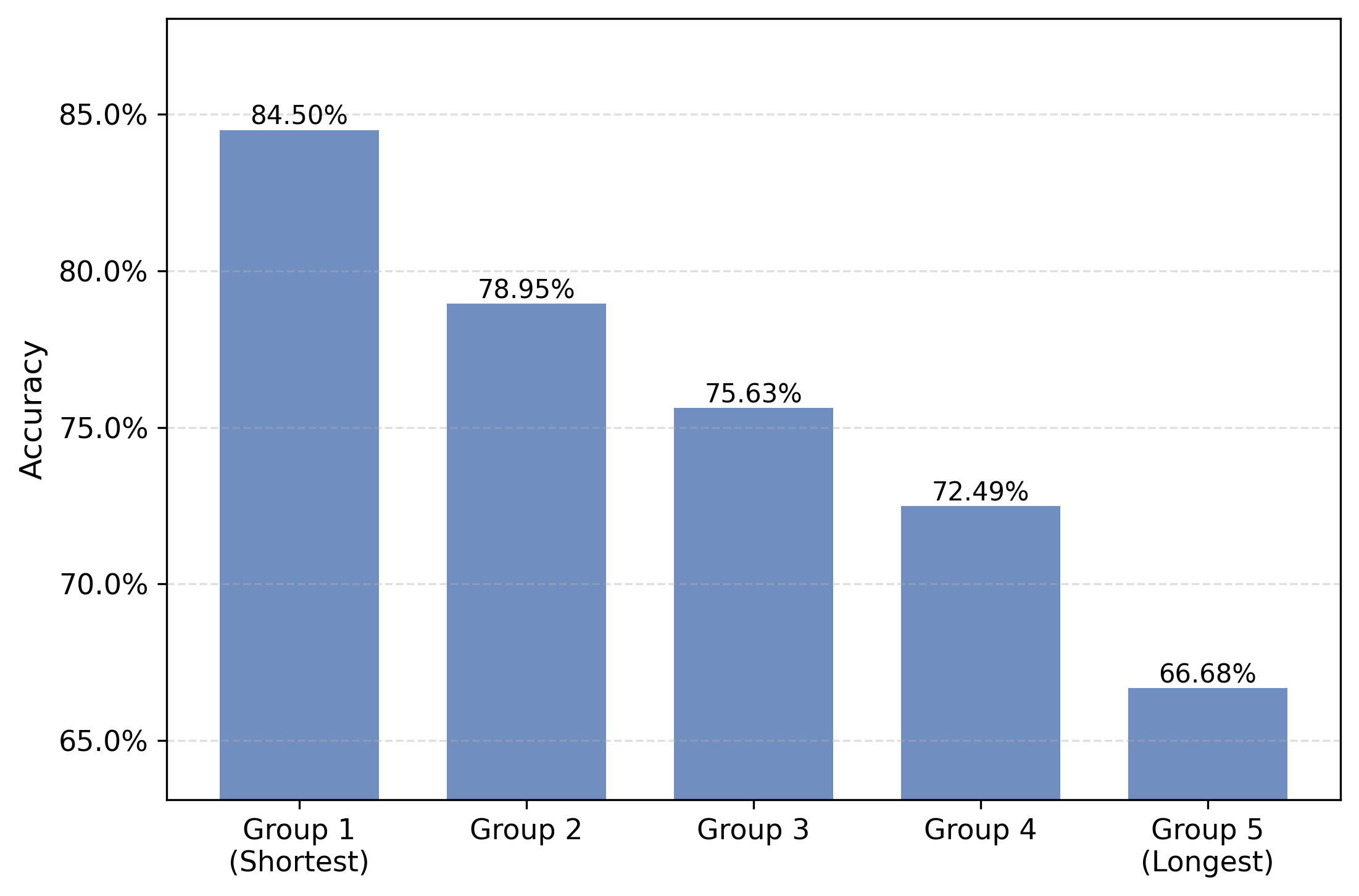} % Replace with your image file
        % \label{fig:subplot2}
        }
    \caption{Comparison of response length \textit{w.r.t.} accuracy in reasoning LLMs.}
    \label{fig:model_len}
\end{figure}

To analyze this, we conducted the following experiment. We compared response lengths and accuracies across various problems. We sorted the responses length in ascending order and categorized the responses into five groups and then calculated the mean accuracy for each group, where accuracy indicates whether the model’s final answer was correct. The results, detailed in Figure  \ref{fig:model_len} reveal an intriguing trend: the shortest responses group (Group 1) achieve the highest accuracy, with performance declining as response length increases. This finding contrasts sharply with observations in mathematical reasoning tasks. The data consistently indicate that shorter responses outperform longer ones in this context. We hypothesize that this may reflect “\textit{overthinking}” by the model, where excessive elaboration introduces redundant steps or errors, undermining the final answer. In contrast to mathematical reasoning, where problems often require multi-step deductions, the tasks in our study may favor direct inference or concise solutions.

\textbf{\textsc{Takeaway} III}: Contrary to trends in mathematical reasoning tasks, shorter responses correlate with improved performance in recommendation tasks.

\begin{figure}[!t]
    \centering
    % First subplot
    \subfigure[BookCrossing]{
        \includegraphics[width=0.31\columnwidth]{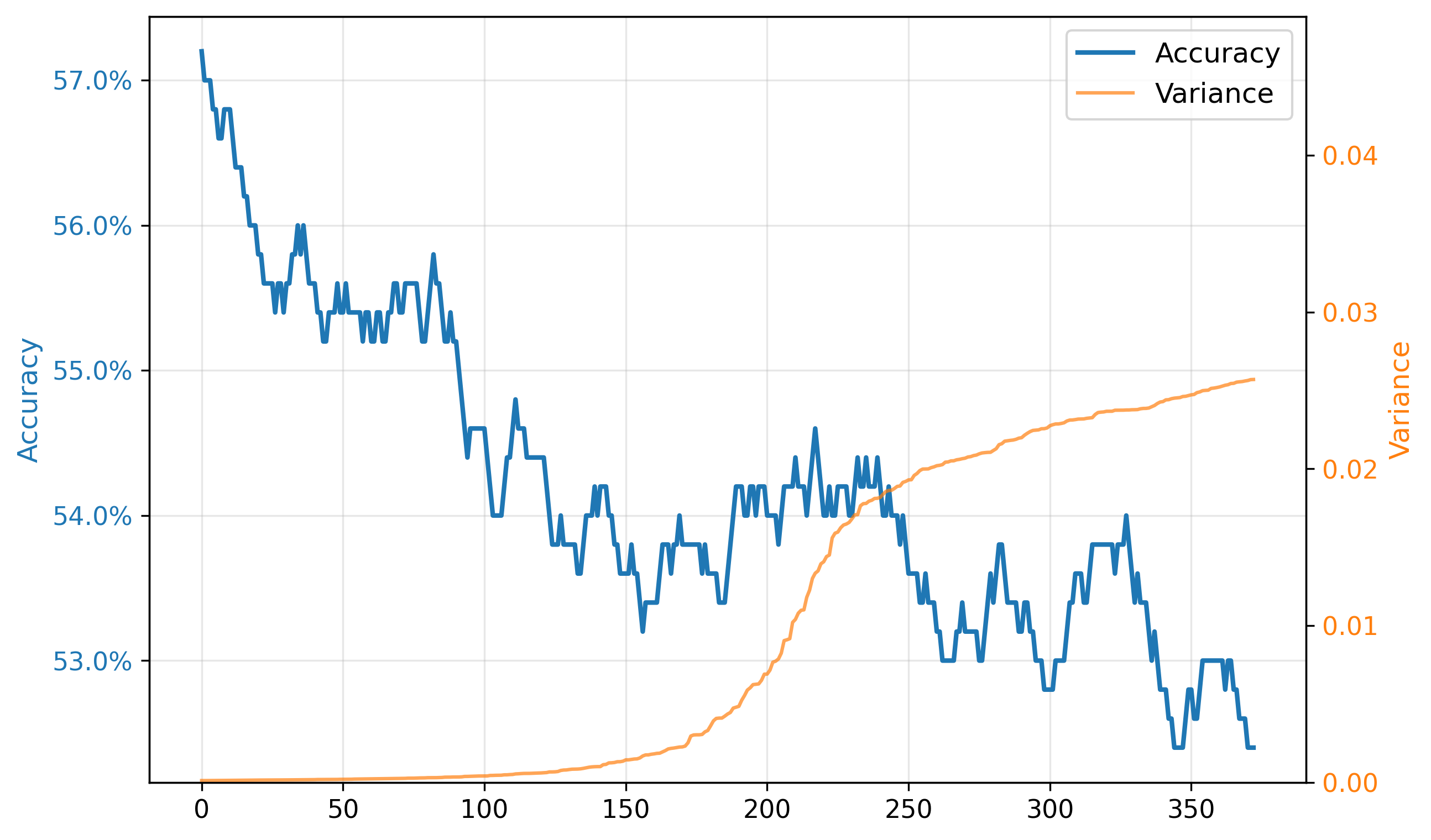} % Replace with your image file
    }
    \hfill
    % Second subplot
    \subfigure[MovieLens]{
        \includegraphics[width=0.31\columnwidth]{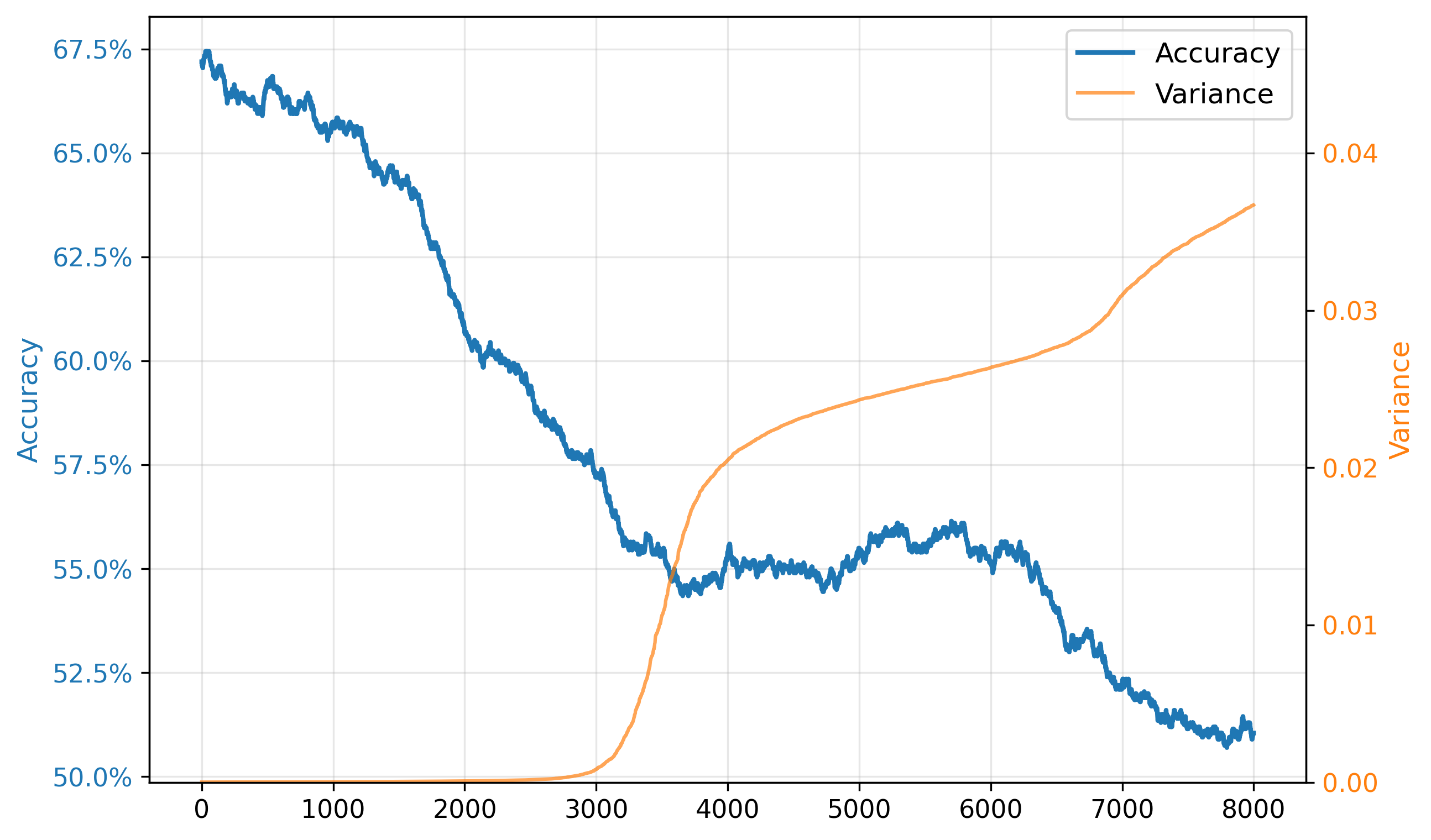} % Replace with your image file
        % \label{fig:subplot2}
        }
    \hfill
    % Second subplot
    \subfigure[Amazon]{
        \includegraphics[width=0.31\columnwidth]{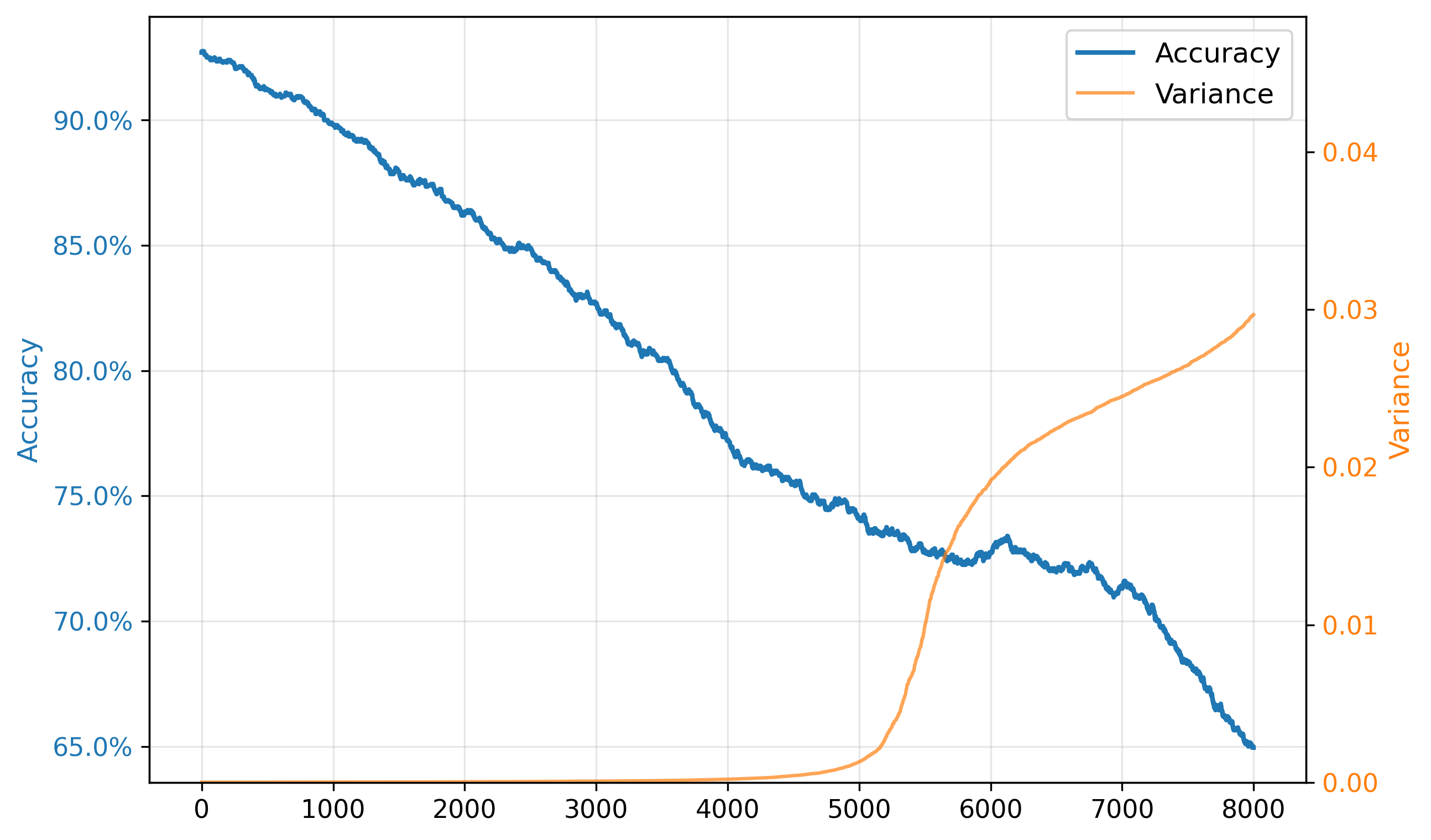} % Replace with your image file
        % \label{fig:subplot2}
        }
    \caption{Comparison of response consistency \textit{w.r.t.} accuracy in reasoning LLMs.}
    \label{fig:Model_cons}
\end{figure}

\textbf{Analysis of Response Consistency in Reasoning LLM.}
We further analyze the response consistency in reasoning LLMs. To this end, we calculate the variance of all responses for each sample, denoted as \( \text{var}(\mathcal{R}) \), where \( \mathcal{R}=\{r_1,...,r_n \} \) represents the set of generated responses. Additionally, we evaluate performance within a sliding window approach.
We repeat the experiment five times, sorting the indices of the responses by their variance for each sample. The sorted data is then divided into distinct windows, and we compute the average accuracy for each window to analyze performance trends. For the BookCrossing dataset, we set the window size to 500 samples, while for the MovieLens and Amazon datasets, we use a window size of 1,000 samples.
As depicted in Figure \ref{fig:Model_cons}, our results indicate a strong correlation between response consistency (\textit{i.e.}, lower variance) and superior performance metrics. This suggests that models generating more consistent responses are likely to exhibit enhanced reliability and effectiveness in reasoning tasks.

\textbf{\textsc{Takeaway} IV}: More consistent responses are generally associated with improved results.

% \textbf{Analysis of model scaling LLM}

% We test with larger reasoning LLM, i.e., 

% \textbf{\textsc{Takeaway} V}: larger model may not perform very well compared with smaller model

% we propose shortest of n, and consistent selction method, which is to only select answer when N response are all consistant. 

% Un tuned reasoning LLM could compare comparable with tuned LLM.

% \textbf{\textsc{Takeaway} \MakeUppercase{\romannumeral 3}: Test time computing techinque (e.g., shortest of n, mahorty voting) could enhance the performance, with similar to .}

\section{Methodology}

\begin{figure}[t]
\centering
\includegraphics[width=0.56\textwidth]{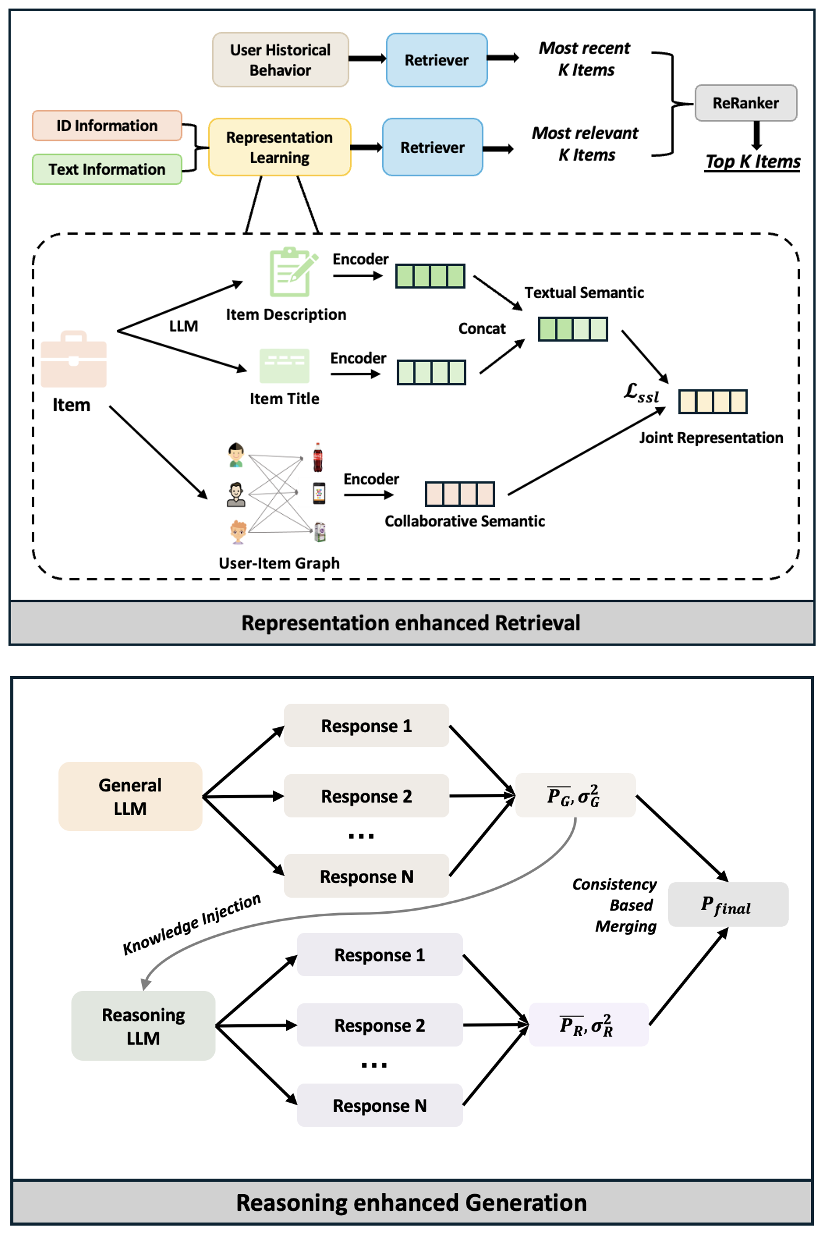}
    \caption{\model with representation learning enhanced retrieval and reasoning enhanced generation.}
\label{fig:main}
% \vspace{-3ex}
\end{figure}

\subsection{Framework Pipeline}

% In this paper, we focus on the click-through rate (CTR) prediction \cite{lin2024rella}. 
The pipeline of the developed framework is illustrated in Figure~\ref{fig:main}.
The \model encompasses both the retrieval and generation processes.
In the retrieval process, we first learn a joint representation of users and items, allowing us to retrieve the most relevant items in semantic space. These items are then fused with the most recent items by a reranker and incorporated into the prompts. The constructed prompts can be used solely for inference or to train a more effective model through instruction tuning (IT).

For the generation phase, the base LLM responds to the prompt for inference, with the option to use either a standard or customized model. We investigate adapting reasoning LLMs for recommendation tasks by first evaluating their performance in this context. Subsequently, we propose a technique to integrate these models into existing systems.
% instruction-tuned
% \lsc{Need polish this part}

% \begin{figure}[t]
% \centering
% \includegraphics[width=0.43\textwidth]{text.png}
% \vspace{-2ex}
%     \caption{Comparison of textual descriptions with fixed template (upper) and automatic generation (blow).}
% \label{fig:text_desc}
% \vspace{-3ex}
% \end{figure}

\subsection{ Representation Learning enhanced Retrieval}
\label{sec:retrieval}

To learn better item embeddings\footnote{We interchangeably use the representation and embedding to denote the extracted item feature considering the habits in deep learning and information retrieval domains.} for reliable retrieval, we propose to integrate both the text embedding from textual description and collaborative embedding from user-item interaction, as well as the joint representation through self-supervised training.

\subsubsection{Textual Representation Learning}

% \subsubsection{Item-side Textual Information Enrichment}

\begin{figure}[t]
\centering
\includegraphics[width=0.63\textwidth]{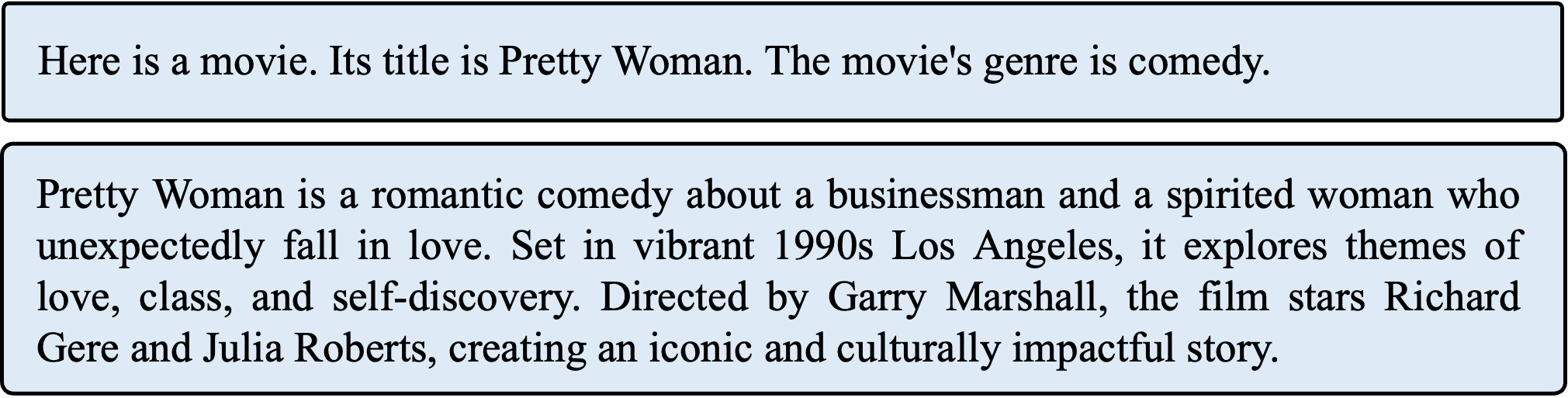}
    \caption{Comparison of textual descriptions with fixed template (upper) and automatic generation (blow).}
\label{fig:text_desc}
\end{figure}

In previous work \cite{lin2024rella}, only the fixed text template with basic information such as item title was utilized to extract textual information. However, we argue that relying solely on the fixed text format is inadequate, as it may not capture sufficient semantic depth, \textit{e.g.}, two distinct and irrelevant items may probably have similar names. To enhance this, we take advantage of the LLMs to generate a more comprehensive and detailed description containing the key attributes of the item (\textit{e.g.}, Figure~\ref{fig:text_desc}), which can be denoted as
\begin{equation}
    t^{i}_{\text{desc}} = \text{LLM}(b^{i}|p),
\end{equation}
where $b^{i}$ is the basic information of the $i$-th item and the $p$ is the template for prompting the LLMs.
Subsequently, we derive textual embeddings by feeding the text into LLMs and taking the hidden representation as in \cite{lin2024rella}, represented as 
\begin{equation}
    \mathbf{e}_{\text{desc}}^i = {{\text{LLM}}}_{emb}(t^{i}_{\text{desc}}).
    \label{eq:llm_emb}
\end{equation}
Since the plain embedding of item title $e_{\text{title}}^i$ could also be useful, we aim to directly concatenate these two kinds of embeddings to obtain the final textual representations, denoted by
\begin{equation}
    \mathbf{e}_{\text{text}}^i = [\mathbf{e}_{\text{title}}^i \| \mathbf{e}_{\text{desc}}^i].
    \label{eq:llm_emb}
\end{equation}
It is worth noting that those textual embeddings are reusable after being extracted and they already contain affinity information attributed to the rich knowledge of LLMs.

\subsubsection{Collaborative Representation Learning}

% \subsubsection{Item-side Collaborative Information}
A notable shortcoming of previous LLM-based approaches is their failure to incorporate collaborative information, which is directed learned from the user-item interaction records and thus can be complementary to the text embeddings. To this end, we utilize conventional recommendation models to extract collaborative semantics, denoted as
\begin{equation}
    \{\mathbf{e}_{\text{colla}}^i\}_{i=1}^{n} = \text{RecModel}(\{(u, i)\in \mathcal{V}\}),
\end{equation}
where $n$ is the total number of items and $\mathcal{V}$ is the interaction history.

\subsubsection{Joint Representation Learning}

A straightforward approach for integrating above two representations is to directly concatenate the textual and collaborative representations. However, since these representations may not be on the same dimension and scale, this might not be the best choice. Inspired by the success of contrastive learning in aligning different views in recommendations \cite{zou2022multi}, we employ a self-supervised learning technique to effectively align the textual and collaborative representations.
Specifically, we adopt a simple two-layer MLP as the projector for mapping the original text embedding space into a lower feature space and use the following self-supervised training objective
\begin{equation} \footnotesize
    \mathcal{L}_{ssl}^{}=-\mathbb{E} \left\{\log \left[\frac{f\left(\mathbf{e}_{\text{text}}^i, \mathbf{e}_{\text{colla}}^i\right)}{\sum_{v \in \mathcal{V}} f\left(\mathbf{e}_{\text{text}}^i, \mathbf{e}_{\text{colla}}^v\right)}\right] + \log \left[ \frac{f\left(\mathbf{e}_{\text{colla}}^i, \mathbf{e}_{\text{text}}^i\right)}{\sum_{v \in \mathcal{V}} f\left(\mathbf{e}_{\text{colla}}^i, \mathbf{e}_{\text{text}}^v\right)}\right]\right\},
    \label{eq:loss_ssl}
\end{equation}
where $f\left(\mathbf{e}_{\text{text}}^i, \mathbf{e}_{\text{colla}}^v\right)=exp(sim(\text{MLP}(\mathbf{e}_{\text{text}}^i), \mathbf{e}_{\text{colla}}^v))$ and $sim(\cdot)$ is the cosine similarity function. After the joint representation learning, we can get the aligned embedding for each item $i$ as
\begin{equation}
    \mathbf{e}_{\text{ssl}}^i = \text{MLP}(\mathbf{e}_{\text{text}}^i).
\end{equation}

\subsubsection{Embedding Mixture}
Instead of retrieval using different embeddings separately, we find that integrating those embeddings before retrieval can present better performance, therefore we directly concat them after magnitude normalization
\begin{equation}
    \mathbf{e}_{\text{item}} = [\mathbf{\bar{e}}_{\text{text}}||\mathbf{\bar{e}}_{\text{colla}}||\mathbf{\bar{e}}_{\text{ssl}}],
\end{equation}
where $\mathbf{\bar{e}} := \mathbf{{e}}/\|\mathbf{{e}}\|$. With the final item embeddings, we can retrieve the most relevant items to the target item by simply comparing the dot-production for downstream recommendation tasks.

% \subsection{User-side Collaborative Information Incorporation}

% \subsection{Data Augmentation}

% Data sparsity is a longstanding problem in recommender systems. In some cases, user history may not be relevant to the target item, potentially due to this sparsity. However, there may be instances where similar users have records that are more aligned with the target items. In such cases, we believe that incorporating information from these similar users can enhance model performance. Specifically, we aim to include similar user records in the prompt to enrich the information provided. The prompt template is as follows:

\subsubsection{Prompt Construction}
To form a prompt message that LLMs can understand, we use a similar template as in \cite{lin2024rella} by filling the user profile, listing the relevant behavior history and instructing the model to give a prediction. We also observed that the pre-trained base LLMs may perform poorly in instruction following. Therefore, we collect a small amount of training data for instruction tuning, where the prompts are constructed with similarity-based retrieval and a \textit{data augmentation} technique is also employed by re-arranging the retrieved sequence according to the timestamp to reduce the impact of item order.

% We observed that the pre-trained base LLMs may perform poorly in the recommendation tasks as the new input and output format is not well aligned. To this end, we collect a small amount training data for supervised fine-tuning (Instruction Tuning). The training data is constructed by embedding-based history retrieval without re-ranking while the inference process can be boosted by re-ranking technique. Further, we re-rank the retrieved sequence according to the timestamps as a kind of data augmentation.

\subsubsection{Reranker}
Since we can retrieve the most recent $K$ items as well as the most relevant $K$ items, relying solely on one of them may not be the optimal choice. During the inference stage, we further innovatively design a reranker to merge these two different channels. The reranker can be either learning-based or rule-based; in this case, we utilize a heuristic rule-based reranker. For each item, we assign a channel score $\text{S}_{c}$ and a position score $\text{S}_{pos}$. We assign the channel score as $\gamma$ and $(1-\gamma)$ for embedding-based and time-based channel, respectively. The position score is inversely proportional to the position in the original sequence, \textit{i.e.}, $\{1, \frac{1}{2^{\beta}}, ..., \frac{1}{K^{\beta}}\}$. The hyper-parameters $\gamma$ and $\beta$ are tunable. The total score for each item is calculated as the production of these two scores 
\begin{equation}
    \text{Score}^i = \text{S}^{i}_{c} * \text{S}^{i}_{pos}.
\end{equation}
By taking the items with top-$K$ scores, we can obtain a refined retrieval result to maximize the prediction performance.

% \subsection{Evaluation Metric}
% We use the logits of the positions of "Yes" and "No" to compute the relative probability of the CTR prediction...

\subsection{ Reasoning enhanced Generation}

Based on the insights in Sec. \ref{sec:analysis}, we propose a knowledge-injected prompting method and a consistency-based merging technique to adapt a reasoning LLM with a general-purpose LLM, resulting in improved performance.

\subsubsection{Aggregate Reasoning and Tuned LLM Synergy}
\label{sec:fusion}

The experiment result in Sec. \ref{sec:analysis} reveals a critical dichotomy: while vanilla reasoning LLMs demonstrate superior structured reasoning capabilities compared to general-purpose LLMs, they underperform domain-tuned LLMs with a large gap \cite{xu2025rallrec}. This presents a fundamental challenge: how to enhance reasoning LLMs effectively?
Supervised fine-tuning (SFT) is a standard approach to align LLMs with domain-specific tasks \cite{luo2024recranker}. Fine-tuning general LLMs requires ground-truth labels.
% bypassing the need for high-quality labeled data. 
However, fine-tuning reasoning LLMs is more challenging: user preferences are highly subjective, making it difficult to craft gold-standard reasoning paths for guidance. Additionally, SFT can be resource-intensive, posing practical limitations.
To solve this, 
our key innovation stems from two mechanisms: \textit{(i)} knowledge-injected prompting, which enriches reasoning LLMs with domain knowledge; and \textit{(ii)} consistency-based merging, which combines reasoning and general LLMs to aggregate their strengths, boosting performance effectively.

\subsubsection{Knowledge-injected prompting}
We propose a novel prompt augmentation strategy that transfers knowledge from the recommendation expert (\textit{e.g}, tuned LLM $\mathcal{M}_G$) to the reasoning LLM ($\mathcal{M}_R$). For input query $x$, we first extract $\mathcal{M}_G$'s prediction $k_G = \mathcal{M}_G(x)$ through its output layer. These predictions are then projected into natural language space and injected into the original query $x$.
% via a lightweight adapter:
% \begin{equation}
% k_G = \text{Proj}_\theta(e_G) \in \mathbb{R}^d
% \end{equation}
Then the knowledge-enhanced prompt for $\mathcal{M}_R$ becomes:

\begin{equation}
p_{\text{aug}} = [\underbrace{\text{[Task Instruction]}}_{{\text{Base Prompt}}}; 
\underbrace{x}_{\text{Input}};\underbrace{k_G}_{\text{Injected Knowledge}}]
\end{equation}

This allows $\mathcal{M}_R$ to take advantage of $\mathcal{M}_G$'s learned patterns while maintaining its intrinsic reasoning capabilities. 
We use the prompt "Another one think the answer might be [Yes/No]" as the injected knowledge for our task.
% The projection network $\theta$ is trained using contrastive learning to maximize mutual information between $k_G$ and $\mathcal{M}_G$'s predictions.

\subsubsection{Consistency-based merging}
Result-level merging across different models offers a simple yet effective approach to integrate their predictions. Drawing from findings in Sec. 3.3.1, more consistent responses correlate with improved outcomes, which we interpret as a measure of model {confidence}. Consequently, we propose assigning higher weights to more confident predictions during merging, enhancing overall performance.

Let $\mathcal{M}_R$ generate $K$ reasoning traces using the augmented prompt, producing mean prediction $\bar{P}_R$ and variance $\sigma^2_R$. The tuned LLM $\mathcal{M}_G$ provides prediction $\bar{P}_G$ with variance $\sigma^2_G$ estimation. Our fusion mechanism combines these through consistency-calibrated weighting:

\begin{equation}
P_{\text{final}} = \frac{\alpha \cdot \frac{\bar{P}_R}{\sigma_R^2 + \epsilon} + \frac{\bar{P}_G}{\sigma_G^2 + \epsilon}}{\alpha \cdot (\sigma_R^2 + \epsilon)^{-1} + (\sigma_G^2 + \epsilon)^{-1}}
\end{equation}

where $\alpha$ is a hyperparameter. The $\epsilon$ term is a small number that ensures numerical stability.

\section{Experiment}

In this section, we assess the performance of our framework and aim to answer the following research questions:

\begin{itemize}[left=0em]
    \item \textbf{RQ1:} How does our proposed \model framework compare with both the conventional recommendation models and the state-of-the-art LLM-based RAG recommendation methods?  
    \item \textbf{RQ2:} Do the designed components of our model function effectively?  
    \item \textbf{RQ3:} How do different hyper-parameters affect the final recommendation performance?  
\end{itemize}

\begin{table*}[t]
\caption{
% The performance of different models in \emph{zero-shot}, \emph{full-shot} and \emph{few-shot} settings. 
% In \emph{full-shot} setting, the baselines are trained on the entire training set. 
% In \emph{few-shot} setting, the number of training shots $N$ is selected from $\{256 (<1\%), 1024(<10\%)\}$ on BookCrossing dataset, and $\{8192 (<1\%), 65536 (<10\%)\}$ on MovieLens-1M and MovieLens-25M datasets. 
% The user behavior sequence length $K$ is set to 60 on BookCrossing and 30 on both MovieLens-1M and MovieLens-25M. 
The performance of different models in default settings. The best results are highlighted in boldface. 
}
% \vspace{-8pt}
\label{tab:zero & few shot performance}
\resizebox{0.95\textwidth}{!}{
\renewcommand\arraystretch{1.0}
\begin{tabular}{c|c|ccc|ccc|ccc}
\toprule
% \hline

\multicolumn{2}{c|}{\multirow{2}{*}{Model}} & \multicolumn{3}{c|}{BookCrossing} & \multicolumn{3}{c|}{MovieLens} & \multicolumn{3}{c}{Amazon} \\ 
\multicolumn{2}{c|}{} & AUC $\uparrow$  & Log Loss $\downarrow$& ACC $\uparrow$& AUC $\uparrow$ & Log Loss $\downarrow$& ACC $\uparrow$& AUC $\uparrow$ & Log Loss $\downarrow$& ACC $\uparrow$\\ 
   \hline 
   
\multicolumn{1}{c|}{\multirow{4}{*}{ID-based}} & DeepFM & 0.5480&0.8521&0.5212& 0.7184&0.6205&0.6636 &0.6419&0.8281&0.7760\\
\multicolumn{1}{c|}{\multirow{4}{*}{}} & xDeepFM & 0.5541&0.9088&0.5304& 0.7199&0.6210&0.6696&0.6395&0.8055&0.7711 \\
\multicolumn{1}{c|}{\multirow{4}{*}{}} & DCN  & 0.5532&0.9356&0.5189 & 0.7212&0.6164&0.6681&0.6369&0.7873&0.7744 \\
\multicolumn{1}{c|}{\multirow{4}{*}{}} & AutoInt  & 0.5478&0.9854&0.5246& 0.7138&0.6224&0.6613 &0.6424&0.7640&0.7543\\
   \hline

\multicolumn{1}{c|}{\multirow{3}{*}{LLM-based}} & Llama3.1 & 0.5894 & 0.6839 & 0.5418 & 0.5865 & 0.6853 & 0.5591 & 0.7025 & 0.7305 & 0.4719 \\ 
\multicolumn{1}{c|}{\multirow{4}{*}{}} & ReLLa & 0.7125 & 0.6458 & 0.6368 & 0.7524 & 0.6182 & 0.6804 & 0.8401 & 0.5074 & 0.8224  \\ 
\multicolumn{1}{c|}{\multirow{4}{*}{}} & Hybrid-Score & 0.7096 & 0.6409 & 0.6334 & 0.7646 & 0.6149 & 0.6843 & 0.8405 & 0.5065 & 0.8256 \\
\hline

\multicolumn{1}{c|}{\multirow{2}{*}{Ours}} & \priormodel & 0.7162 & 0.6365 & 0.6506 & 0.7658 & 0.6140 & 0.6942 & \textbf{0.8416} & 0.5059 & 0.8331 \\ 
\multicolumn{1}{c|}{\multirow{2}{*}{}} 
& \model & \textbf{0.7175} & \textbf{0.6354} & \textbf{0.6518} & \textbf{0.7663} & \textbf{0.6118} & \textbf{0.6948} & 0.8412 & \textbf{0.5036} & \textbf{0.8343} \\ 

% & \textit{p-value} & {0.000869} & {0.00235} & {0.00122} & {0.000003} & {2.051616654292588e-05} & {0.002577} & {0.000139} & {1.96e-05} & {0.00388}\\ 
  
   % \hline  
   \bottomrule          
\end{tabular}
% \vspace{-5ex}
}
\end{table*}

\begin{table*}[t]
\caption{ The performance of different variants of \model. We remove different components of \model to evaluate the contribution of each part to the model. 
The best results are highlighted in boldface. KP refer to knowledge inject prompting and CM refer to consistency based merging.
% given in bold, and the second-best value is underlined. 
}
% \vspace{-6pt}
\label{tab:ablation_train}
\resizebox{\textwidth}{!}{
\renewcommand\arraystretch{1.0}
\begin{tabular}{cc|ccc|ccc|ccc}
\toprule
% \multicolumn{2}{c|}{\multirow{2}{*}{Model}} & \multicolumn{4}{c}{MovieLens-1M} \\ 
% \multicolumn{2}{c|}{} & AUC  & Log Loss & ACC & Rel.Impr\\ 
\multicolumn{2}{c|}{Variants} & \multicolumn{3}{c|}{BookCrossing} & \multicolumn{3}{c|}{MovieLens} & \multicolumn{3}{c}{Amazon} \\ 
\multicolumn{1}{c}{\textit{w/} KP} & \multicolumn{1}{c|}{\textit{w/} CM} & AUC $\uparrow$ & Log Loss $\downarrow$ & ACC $\uparrow$ & AUC $\uparrow$ & Log Loss $\downarrow$ & ACC $\uparrow$ & AUC $\uparrow$ & Log Loss $\downarrow$ & ACC $\uparrow$ \\
\midrule 
\ding{55} & \ding{55} & 0.7141 & 0.6392 & 0.6483 & 0.7641 & 0.6160 & 0.6945 & 0.8397 & 0.5071 & 0.8335 \\
\ding{55} & \ding{51} & 0.7158 & 0.6375 & 0.6506 & 0.7639 & 0.6146 & 0.6944 & 0.8404 & 0.5052 & 0.8332 \\
\ding{51} & \ding{55} & 0.7163 & 0.6363 & 0.6506 & 0.7656 & 0.6127 & 0.6940 & 0.8405 & 0.5044 & 0.8337 \\
\ding{51} & \ding{51} & \textbf{0.7175} & \textbf{0.6354} & \textbf{0.6518} & \textbf{0.7663} & \textbf{0.6118} & \textbf{0.6948} & \textbf{0.8412} & \textbf{0.5036} & \textbf{0.8343} \\

\bottomrule          
\end{tabular}
}
% \vspace{-1ex}
\end{table*}

% \lsc{Introduce more about data processing}

\subsection{Baseline}
We compare our approach with baseline methods, which include both ID-based and LLM-based recommendation systems.
For ID-based methods, we select DeepFM \cite{guo2017deepfm}, xDeepFM \cite{lian2018xdeepfm}, DCN \cite{wang2017deep}, and AutoInt \cite{song2019autoint} as our baseline models. We utilize Llama3.1-8B-Instruct \cite{dubey2024llama} as the base model and LightGCN \cite{He2020LightGCN} to learn collaborative embeddings in our comparisons.
For LLM-based methods, we consider ReLLa \cite{lin2024rella} and a Hybrid-Score based retrieval as in \cite{zeng2024federated}. 
\priormodel includes solely on representation learning enhanced retrieval, while \model integrates both representation learning enhanced retrieval and reasoning enhanced generation.
By default, we apply the LoRA method 
\cite{hu2022lora} and 8-bit quantization to conduct instruction-tuning as in \cite{lin2024rella} and the maximum length of history is $K=30$. Similar to ReLLa \cite{lin2024rella}, we collect the user history sequence before the latest item and the ratings to construct the prompting message and ground-truth. For the reranker in our method, we search the $\gamma$ over $\{\frac{1}{2}, \frac{2}{3}, \frac{4}{5}\}$ and fix $\beta=1$ in the experiments.
We set $\alpha$ to 0.1 and $\epsilon$ to e$^{-3}$ for all experiments unless specifically specified.

\subsection{Main Result}

% \textbf{Sequential Behavior Comprehension.} 
From the numerical results presented in Table~\ref{tab:zero & few shot performance}, several noteworthy observations emerge. Firstly, the vanilla ID-based methods generally underperform LLM-based methods, demonstrating that LLMs can better leverage textual and historical information for preference understanding. 
% Notably, these conventional methods fail to capture fine-grained semantic associations from item descriptions and user history presented in a natural language format, which limits their performance.
Secondly, among LLM-based baselines, ReLLa effectively incorporates a retrieval-augmented approach but relies predominantly on simple textual semantics for item retrieval. Hybrid-Score, which considers both ID-based and textual features, also improves over the zero-shot LLM setting (Llama3.1). However, both ReLLa and Hybrid-Score still fail to fully leverage the rich collaborative semantics and the alignment between textual and collaborative embeddings. In contrast, \priormodel and \model consistently achieve the best results across all three datasets, outperforming both ID-based and LLM-based baselines. 
% For instance, on the BookCrossing dataset, \model improves the AUC by more than 0.27 over Llama3.1 and about 0.26 over DeepFM. On the larger MovieLens and Amazon datasets, \model also achieves significant performance gains. 
The improvements are statistically significant with $p$-values less than 0.01, emphasizing the robustness of our approach.

\subsection{Ablation Study}

To assess the contributions of key components in \model, we conduct ablation studies by systematically removing Knowledge injected Prompting (KP) and Consistency based Merging (CM). Results are shown in Table~\ref{tab:ablation_train}, and the best performance is highlighted in boldface.

We observe a noticeable decline in AUC and an increase in log loss when removing KP, underscoring KP’s critical role in enhancing the reasoning LLM’s contextual understanding. It consistently weakens performance across all datasets without injecting domain knowledge into the reasoning LLM’s prompts.
Additionally, 
when excluding CM, which aligns predictions from reasoning LLM and the tuned LLM using consistency metrics, also reduces performance, highlighting CM’s role in stabilizing the fusion process.
The full model, incorporating both KP and CM, consistently outperforms ablated variants across all metrics and datasets, achieving the highest AUC, lowest log loss, and best accuracy. Removing both components yields the weakest results, confirming their complementary strengths: KP boosts reasoning capabilities, while CM ensures effective prediction alignment.

\begin{figure}[!t]
    \centering
    % First subplot
    \subfigure[BookCrossing]{
        \includegraphics[width=0.31\columnwidth]{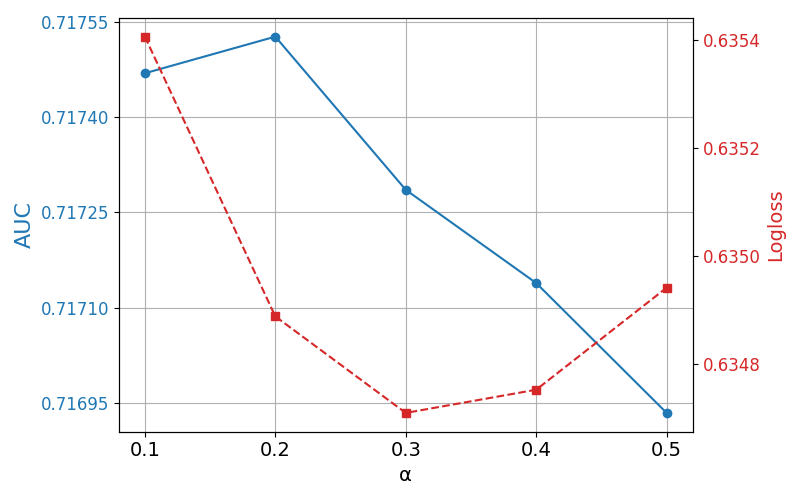} % Replace with your image file
    }
    \hfill
    % Second subplot
    \subfigure[MovieLens]{
        \includegraphics[width=0.31\columnwidth]{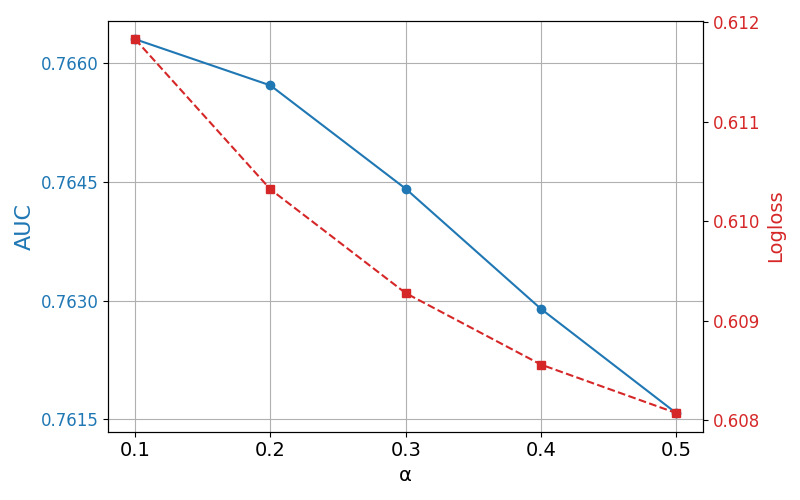} % Replace with your image file
        % \label{fig:subplot2}
        }
    \hfill
    % Second subplot
    \subfigure[Amazon]{
\includegraphics[width=0.31\columnwidth]{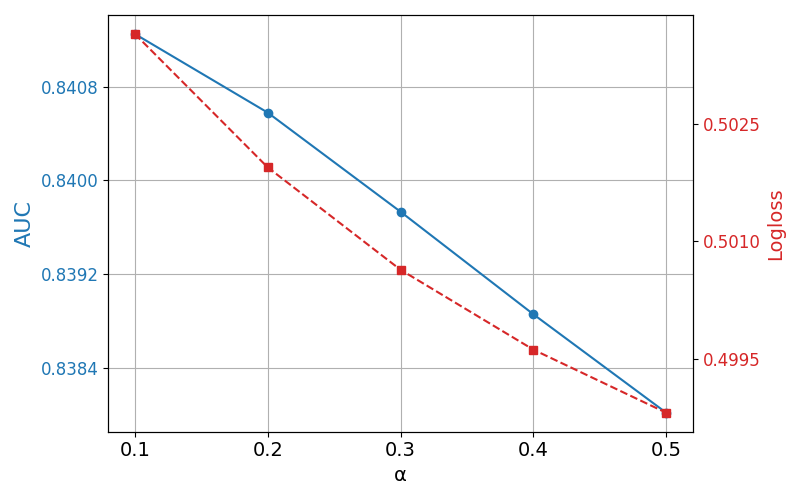} % Replace with your image file
        % \label{fig:subplot2}
        }
    \caption{Impact of \(\alpha\) on reasoning LLMs performance across three datasets.}
    \label{fig:alpha_study}
\end{figure}

\begin{figure}[!t]
    \centering
    % First subplot
    \subfigure[BookCrossing]{
        \includegraphics[width=0.31\columnwidth]{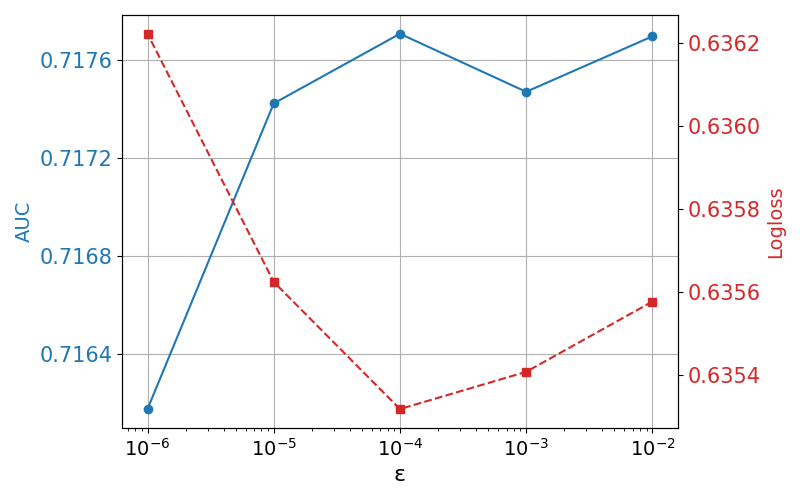} % Replace with your image file
    }
    \hfill
    % Second subplot
    \subfigure[MovieLens]{
        \includegraphics[width=0.31\columnwidth]{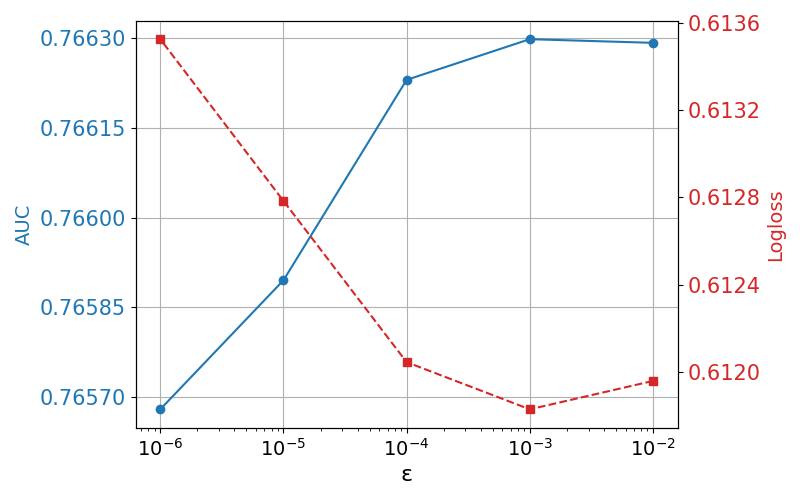} % Replace with your image file
        % \label{fig:subplot2}
        }
    \hfill
    % Second subplot
    \subfigure[Amazon]{
\includegraphics[width=0.31\columnwidth]{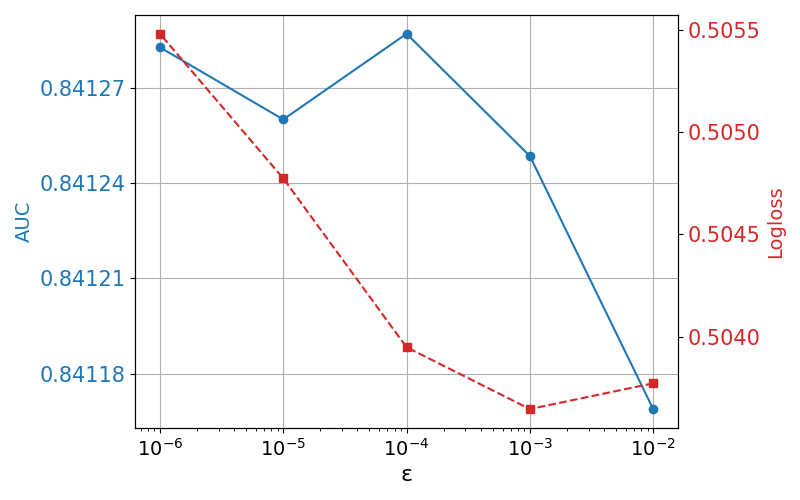} % Replace with your image file
        % \label{fig:subplot2}
        }
    \caption{Impact of \(\epsilon\) on reasoning LLMs performance across three datasets.}
    \label{fig:epsilon_study}
\end{figure}

\subsection{Analysis of Retrieval}
In this section, we evaluate the retrieval mechanisms of the proposed method, focusing solely on the representation learning and reranking components without incorporating reasoning LLMs. By isolating these elements, we assess their standalone effectiveness in enhancing item retrieval for recommendation tasks. Experiments compare embedding strategies and reranking approaches across datasets, providing insights into their contributions to overall performance.

\begin{figure}[t]
\centering
\includegraphics[width=0.69\textwidth]{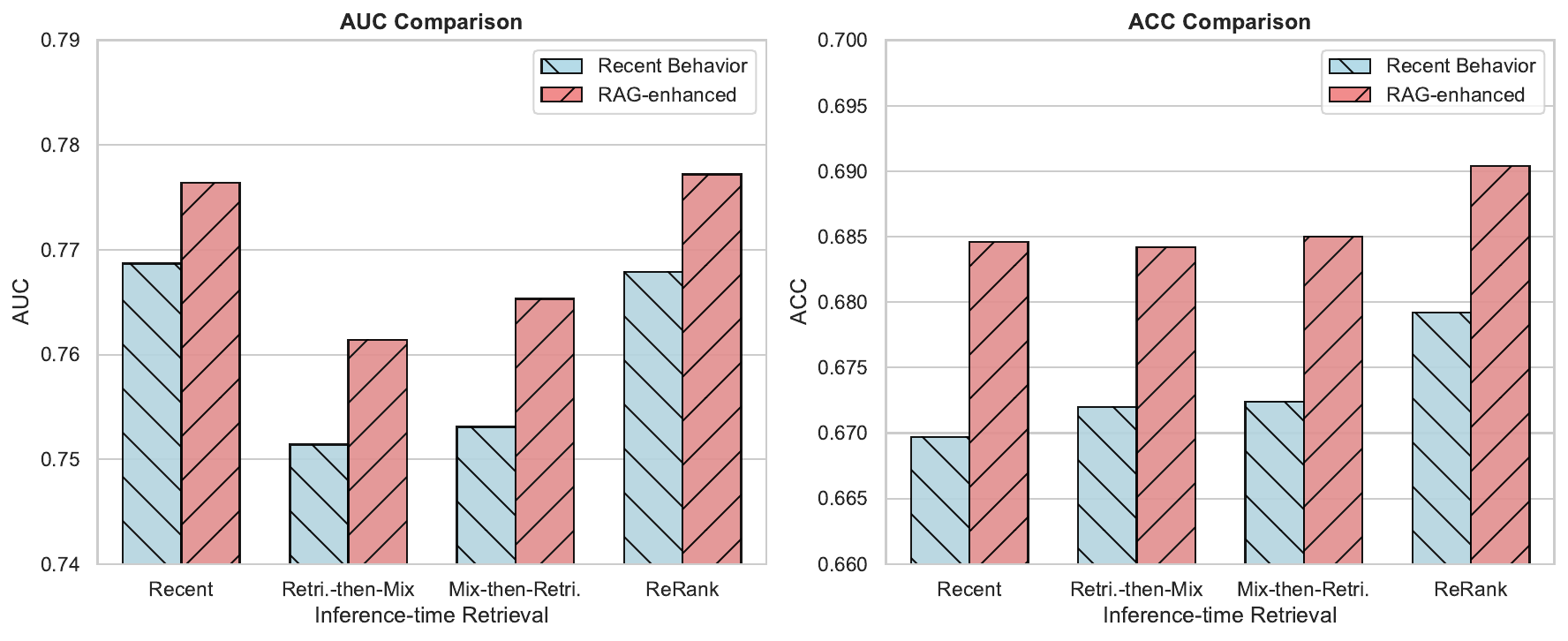}
% \vspace{-5ex}
    \caption{Comparison of fine-tuning and inference settings.}
\label{fig:ablation_prompt}
% \vspace{-3ex}
\end{figure}

\begin{table*}[t]
\caption{The comparison of different embeddings used for historic behavior retrieval during inference. For fair comparisons, the model is instruction-tuned using the RAG-enhanced training data, while the inference prompt is constructed based on the embedding similarity without re-ranking. The best results are highlighted in boldface.
}
% \vspace{-8pt}
\label{tab:ablation_emb}
\resizebox{1.0\textwidth}{!}{
\renewcommand\arraystretch{1.0}
\begin{tabular}{c|ccc|ccc|ccc}
\toprule
% \multicolumn{2}{c|}{\multirow{2}{*}{Model}} & \multicolumn{4}{c}{MovieLens-1M} \\ 
% \multicolumn{2}{c|}{} & AUC  & Log Loss & ACC & Rel.Impr\\ 
\multicolumn{1}{c|}{\multirow{2}{*}{Embedding Variant}} & \multicolumn{3}{c|}{BookCrossing} & \multicolumn{3}{c|}{MovieLens} & \multicolumn{3}{c}{Amazon} \\ 
\multicolumn{1}{c|}{} & AUC $\uparrow$  & Log Loss $\downarrow$& ACC $\uparrow$ & AUC $\uparrow$  & Log Loss $\downarrow$& ACC $\uparrow$ & AUC $\uparrow$  & Log Loss $\downarrow$& ACC $\uparrow$ \\ 
   \hline 
% Plain Text & -- & -- & -- & -- & {--} & -- & -- & -- & {--} \\ 
Text-based & 0.7034 & 0.6434 & 0.6426 & 0.7583 & 0.6188 & 0.6798 & 0.8408 & 0.4931 & 0.8222 \\ 
ID-based & 0.7084 & 0.6414 & 0.6357 & 0.7580 & 0.6153 & 0.6867 & 0.8431 & 0.4930 & 0.8244 \\ 
Concat. w/o SSL & 0.7127 & 0.6411 & 0.6391 & 0.7633 & 0.6153 & 0.6828 & 0.8439 & 0.4925 & 0.8244 \\ 
Concat. w/ SSL & \textbf{0.7141} & \textbf{0.6413} & \textbf{0.6471} & \textbf{0.7653} & \textbf{0.6144} & \textbf{0.6850} & \textbf{0.8442} & \textbf{0.4924} & \textbf{0.8269} \\  
\bottomrule          
\end{tabular}
}
% \vspace{-3ex}
\end{table*}

\subsubsection{Reranking and Retrieval Methods} Figure~\ref{fig:ablation_prompt} compares different retrieval and prompt construction approaches on the MovieLens.
% Without fine-tuning, using embedding-based retrieval alone provides limited gains. Once fine-tuned, integrating retrieval with reranking yields consistent improvements. 
We observe the retrieval-then-mix strategy achieves worse performance regarding the AUC metric.
Our reranker, which balances semantic relevance and temporal recency, outperforms both plain recent-history-based prompts and simple hybrid retrieval strategies. These results emphasize the necessity of refining retrieved items through post-processing rather than relying solely on a single retrieval strategy.

\vspace{-1ex}
\subsubsection{Embedding Strategies} In Table~\ref{tab:ablation_emb}, we evaluate different embedding schemes for retrieval. Text-based embeddings, derived from LLM-generated descriptions, yield suboptimal performance compared to ID-based embeddings, which leverage user-item interaction signals more effectively. Concatenating these two representations outperforms either alone, achieving better results by capturing both textual and collaborative insights. Further enhancement is observed when aligning the concatenated embeddings with self-supervised learning, which refines their semantic coherence and boosts performance across datasets. These results highlight the progressive improvement from single-modality embeddings to joint representations.

\subsection{Hyperparameter Studies}

\subsubsection{Study of \(\alpha\)}
The hyperparameter \(\alpha\) balances the contributions of the reasoning LLM and the tuned LLM in the final prediction \(P_{\text{final}}\). We evaluate its impact by varying \(\alpha\) over \([0.1, 0.2, 0.3, 0.4, 0.5]\), shifting influence from \(\mathcal{M}_G\) (smaller \(\alpha\)) to \(\mathcal{M}_R\) (larger \(\alpha\)), while fixing \(\epsilon = 10^{-3}\) to isolate \(\alpha\)'s effect.
The results are shown in Figure~\ref{fig:alpha_study}. We observe distinct trends emerge across different datasets. For BookCrossing, AUC peaks at a moderate \(\alpha\) before a slight decline, log loss improves with increasing \(\alpha\), and accuracy maximizes at a higher \(\alpha\). For Amazon, AUC decreases steadily as \(\alpha\) rises, log loss improves, and accuracy remains stable. In MovieLens, AUC declines gradually, log loss decreases, and accuracy peaks at an intermediate \(\alpha\).
These patterns indicate that smaller \(\alpha\) values favor \(\mathcal{M}_G\), boosting AUC, while larger \(\alpha\) enhances log loss by leveraging \(\mathcal{M}_R\)'s reasoning. Accuracy varies by dataset, reflecting differing sensitivities to reasoning contributions. 
% Given AUC's priority, we set \(\alpha = 0.1\) as the default, ensuring robust performance across all datasets.

\subsubsection{Study of \(\epsilon\)}
The parameter \(\epsilon\) stabilizes prediction weighting in the fusion equation by preventing division-by-zero errors, based on variances (\(\sigma_R^2\) and \(\sigma_G^2\)). We test its effect over \([10^{-2}, 10^{-3}, 10^{-4}, 10^{-5}, 10^{-6}]\), with \(\alpha = 0.1\) fixed.
Figure~\ref{fig:epsilon_study} presents the results. For BookCrossing, performance optimizes at an intermediate \(\epsilon\), with AUC, log loss, and accuracy peaking, though very small \(\epsilon\) values degrade AUC. For Amazon, AUC and log loss improve up to a mid-range \(\epsilon\), with accuracy favoring a larger value. In MovieLens, AUC and log loss stabilize mid-range, while accuracy peaks slightly higher.
These trends suggest that an intermediate \(\epsilon\) balances performance and stability across datasets, avoiding degradation seen at extremes.
% While some metrics favor other settings, \(\epsilon = 10^{-4}\) offers consistency and is adopted as the default.

\begin{table}[!t]
    \caption{The statistics for DeepSeek-Llama generated token per response across all datasets.}
    % \vspace{-8pt}
    \centering
    \resizebox{0.55\textwidth}{!}{
    \renewcommand\arraystretch{1.1}
    \begin{tabular}{c|cccccc}
    \toprule
     Dataset   & BookCrossing & MovieLens & Amazon \\ 
     \midrule
      \# Tokens/Response &687.2&740.3&661.6  \\ 
     \bottomrule
    \end{tabular}
    }
    % \vspace{-5pt}
    \label{tab:tokens}
\end{table}

\begin{figure}[!t]
    \centering
    % First subplot
    \subfigure[BookCrossing]{
        \includegraphics[width=0.28\columnwidth]{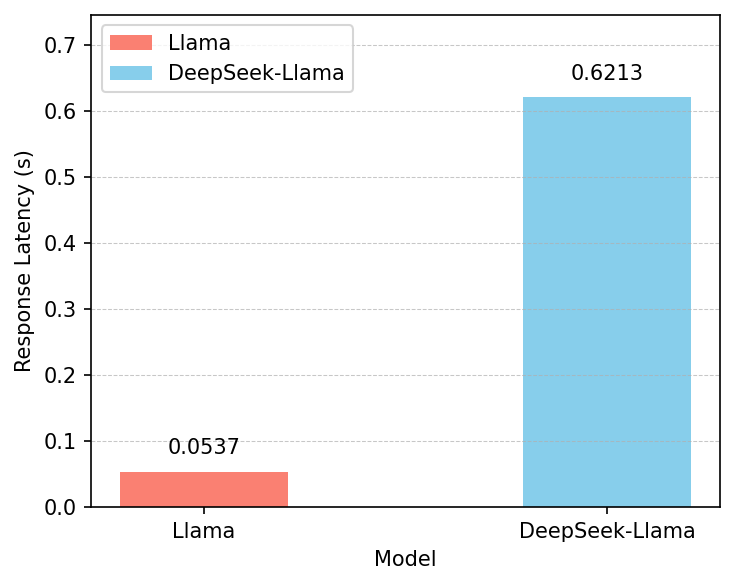} % Replace with your image file
    }
    \hfill
    % Second subplot
    \subfigure[MovieLens]{
        \includegraphics[width=0.28\columnwidth]{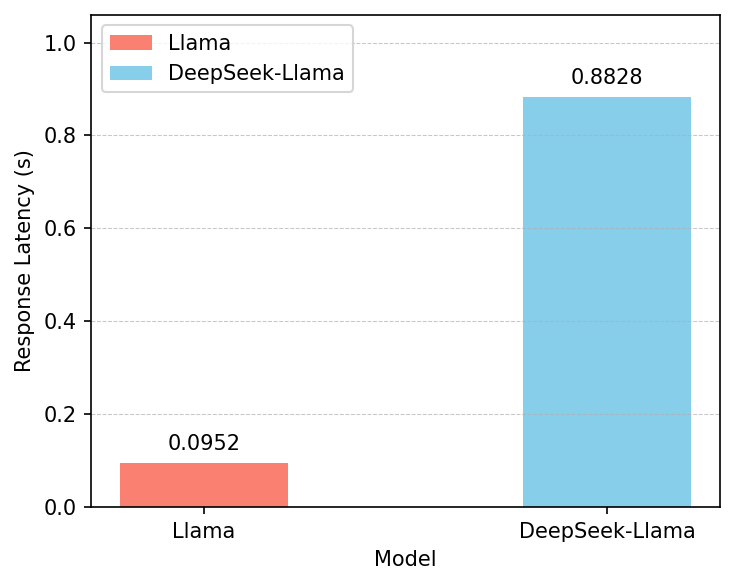} % Replace with your image file
        % \label{fig:subplot2}
        }
    \hfill
    % Second subplot
    \subfigure[Amazon]{
        \includegraphics[width=0.28\columnwidth]{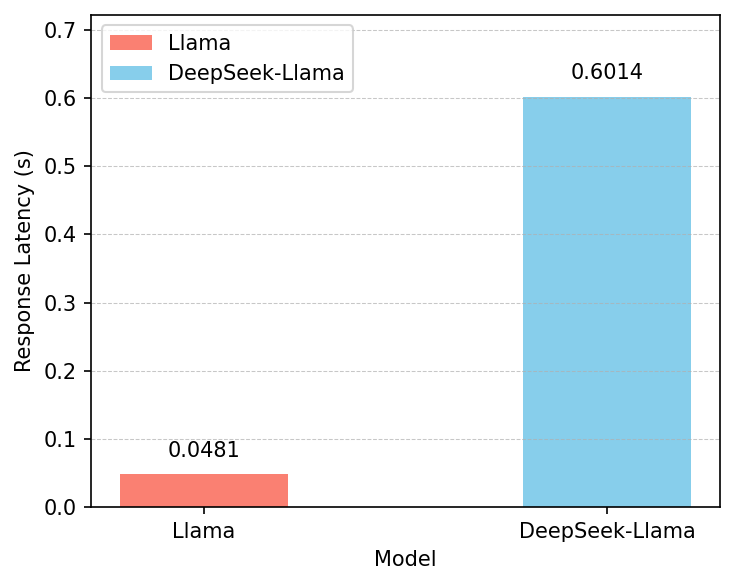} % Replace with your image file
        % \label{fig:subplot2}
        }
    \caption{Comparison of response latency \textit{w.r.t.} Llama and DeepSeek-Llama model across three datasets.}
    \label{fig:model_latency}
\end{figure}

\subsection{Analysis of  Model Efficiency}

Reasoning LLMs, despite their advanced capabilities, exhibit slower performance compared to general LLMs due to the generation of extensive chains of thought. To assess this, we compare the inference latency of a general LLM (Llama-3.1-8B-Instruct, short for Llama) and a reasoning LLM (DeepSeek-R1-Distill-Llama-8B, short for DeepSeek-Llama) across three datasets using the vLLM \cite{kwon2023efficient} framework for acceleration. 
The hardware for the platform includes Intel(R) Xeon(R) Gold 6354 CPU @ 3.00GHz and
NVIDIA 48G A40 GPU.
Latency is measured as the time per response in seconds.
The result is shown in Figure \ref{fig:model_latency}.
We observe the general LLM demonstrates significantly higher efficiency compared to the reasoning LLM across all datasets. On average, the reasoning LLM's inference latency is over 11 times greater than that of the general LLM. This substantial difference is primarily due to the reasoning LLM generating much longer responses, with an average of nearly 700 tokens per response, as detailed in Table~\ref{tab:tokens}.

\section{Conclusion}
% \vspace{-0.5ex}
% In this paper, we introduce a new representation learning framework of item embeddings for LLM-based Recommendation (\model), which improves item description generation and enables joint representation learning of textual and collaborative semantics. Experiments on three datasets demonstrate its capability to retrieve relevant items and improve overall performance. 

We introduce \model, a framework that enhances Retrieval-Augmented LLM recommendations by integrating representation learning and reasoning. It combines textual and collaborative semantics through self-supervised learning and leverages reasoning LLMs for improved performance. Experiments on three datasets show \model outperforms conventional and state-of-the-art methods. Our findings reveal that reasoning LLMs excel in recommendations, retrieval augmentation boosts reasoning, and response consistency correlates with better result. 
Future work could explore fine-tuning reasoning LLMs to deepen their task-specific capabilities, integrate reinforcement learning to dynamically refine recommendation policies, and investigate hybrid architectures blending reasoning and generative strengths.

\bibliography{ref}

% \section{Appendix / supplemental material}

% Optionally include supplemental material (complete proofs, additional experiments and plots) in appendix.
% All such materials \textbf{SHOULD be included in the main submission.}

%%%%%%%%%%%%%%%%%%%%%%%%%%%%%%%%%%%%%%%%%%%%%%%%%%%%%%%%%%%%

\end{document}